\documentclass[aps,prb,bibnotes,singlecolumn,showpacs,preprintnumbers,amsmath,amssymb,superscriptaddress,floatfix]{revtex4-2}

\usepackage[T1]{fontenc}
\usepackage[ansinew]{inputenc}
\usepackage{float}
\usepackage{graphics}
\usepackage{dcolumn}
\usepackage{bm}
\usepackage{amsthm}
\usepackage{amsmath}
\usepackage{amssymb}
\usepackage{diagbox}
\usepackage{hyperref}
\usepackage{url}
\usepackage{xcolor}
\usepackage{verbatim}
\usepackage{minted}
\usepackage{tcolorbox}
\usepackage{listings}

\begin{document}

\title{NuMagSANS: a GPU-accelerated open-source software package for the generic computation of nuclear and magnetic small-angle neutron scattering observables of complex systems}

\author{Michael P.\ Adams}\email[Electronic address: ]{michael.adams@uni.lu}
\affiliation{Department of Physics and Materials Science, University of Luxembourg, 162A~Avenue de la Faiencerie, L-1511 Luxembourg, Grand Duchy of Luxembourg}

\author{Andreas Michels}\email[Electronic address: ]{andreas.michels@uni.lu}
\affiliation{Department of Physics and Materials Science, University of Luxembourg, 162A~Avenue de la Faiencerie, L-1511 Luxembourg, Grand Duchy of Luxembourg}

\begin{abstract}
We present \texttt{NuMagSANS}, a GPU-accelerated software package for calculating nuclear and magnetic small-angle neutron scattering (SANS) cross sections and correlation functions. The program allows users to import position-dependent nuclear density and magnetization data, providing a large flexibility for analyzing the scattering signatures of complex systems, particularly magnetic materials. Full rotational control of the sample is supported, allowing a comprehensive exploration of angular-dependent scattering features. \texttt{NuMagSANS} includes a versatile library of approximately 100 response functions that encompass two-dimensional SANS cross sections, correlation functions, and azimuthally averaged quantities. These capabilities allow users to gain detailed insight into the structural and magnetic characteristics of their samples. GPU acceleration ensures rapid computations, even for large data sets, making \texttt{NuMagSANS} a powerful and efficient tool for advanced SANS analysis.
\end{abstract}

\date{\today}

\maketitle


\section{Introduction}

\texttt{NuMagSANS} is a GPU-accelerated software package developed to overcome the computational challenges associated with accurately evaluating nuclear and magnetic small-angle neutron scattering (SANS) observables in structurally and magnetically complex systems. Taking advantage of the parallel processing capabilities of NVIDIA GPUs, \texttt{NuMagSANS} delivers substantial performance improvements compared to conventional CPU-based approaches. This acceleration enables efficient and precise computation of two-dimensional SANS cross sections and correlation functions from user-defined real-space nuclear scattering-length density and magnetization data. This functionality distinguishes \texttt{NuMagSANS} from existing software solutions, including SASfit~\cite{kohlbrecher2022}, the mag2exp software~\cite{Holt2025}, or the generic SAS calculator available within the open-source SasView platform (\url{https://www.sasview.org}).

\texttt{NuMagSANS} is implemented in C++~\cite{stroustrup_cpp} and CUDA~\cite{nvidia_cuda_guide} and is distributed as open-source software under the MIT License at \url{https://github.com/AdamsMP92/NuMagSANS.git}. In addition to the terms of the MIT License, users are kindly requested to cite this article when employing \texttt{NuMagSANS} in scientific publications. For installation, the source code can be downloaded through the above GitHub link and compiled using the NVIDIA CUDA compiler (using the `nvcc' command).

The execution of \texttt{NuMagSANS} requires an NVIDIA GPU and a supported operating system (Linux, macOS, or Windows). Aside from the standard NVIDIA GPU drivers, no external dependencies are required, facilitating a straightforward installation and configuration process.

\texttt{NuMagSANS} was developed during several research projects that aimed to advance the understanding of magnetic SANS. The core code was used in studies to simulate the scattering of nanoparticles with N\'eel surface anisotropy~\cite{adamsjacnum2022,adams2024signature} and to analyze the angular anisotropy of the magnetic SANS cross section of randomly oriented Stoner-Wohlfarth particles~\cite{michaeliucrj2023}. It also enabled detailed investigations of magnetic nanoparticles with pore-type defects~\cite{evelynprb2023}, vortex-type spin structures~\cite{adams2024framework}, and the spin-flip SANS related to the Dzyaloshinskii-Moriya interaction~\cite{sinaga2024neutron}.

Beyond the work presented in the above publications, the current version of \texttt{NuMagSANS} incorporates several important features, including support for nuclear scattering and nuclear-magnetic interference terms. A key strength of \texttt{NuMagSANS} lies in its seamless integration with micromagnetic simulation tools such as \texttt{MuMax3}~\cite{mumax3new,Leliaert_2018}, allowing researchers to directly compute magnetic SANS observables from simulated spin configurations. This capability broadens the scope of the software and enhances its utility to investigate a wide range of magnetic systems.

A particularly promising feature of \texttt{NuMagSANS} is its ability to separate particle form factors from interparticle interference contributions, providing a robust framework for analyzing structure factor effects. This functionality allows researchers to systematically examine how (nuclear and magnetic) interparticle correlations influence scattering cross sections, thereby offering new opportunities to uncover subtle features in complex materials. For example, in studies of skyrmion lattices, isolating the form factor of individual skyrmions and subsequently incorporating the Bragg peaks associated with the skyrmion lattice can yield valuable insights into their mutual interactions. By enabling such detailed analyzes, \texttt{NuMagSANS} provides a versatile and powerful platform for advancing the understanding of intricate scattering phenomena in purely magnetic and hybrid (i.e., nuclear and magnetic) systems.

Four benchmark data sets supporting this work are openly available on Zenodo as part of the \texttt{NuMagSANS} framework~\cite{Adams2025_Example1,Adams2025_Example2,Adams2025_Example3,Adams2026_Example4}. Each data set contains the corresponding magnetic configuration in real space, the \texttt{NuMagSANS} input and output files, and visualization figures. Together, they provide a reproducible benchmark suite for testing and validating SANS simulations within the \texttt{NuMagSANS} framework.

The article is organized as follows. In Section~\ref{sec:NuclearAndMagnetSANS} we provide the expressions for the nuclear and magnetic SANS cross sections and related quantities, such as the pair-distance distribution function and the correlation function. In Section~\ref{sec:ProgramOrgranization} we describe the implementation and  workflow of \texttt{NuMagSANS}. Finally, Section~\ref{examples} features a few example cases.

\section{SANS observables}
\label{sec:NuclearAndMagnetSANS}

\begin{figure}[ht]
\centering
\resizebox{0.70\columnwidth}{!}{\includegraphics{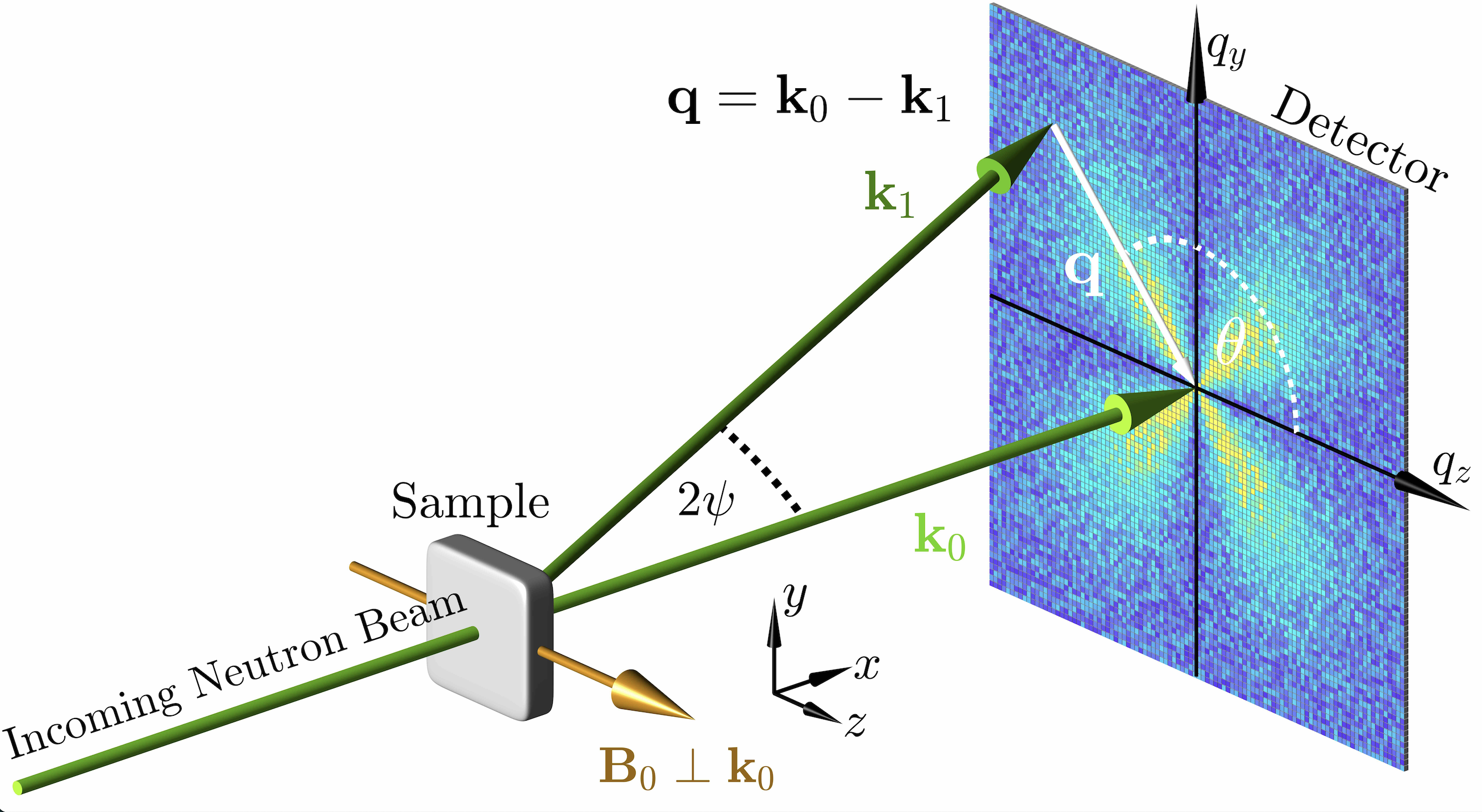}}
\caption{Sketch of the neutron scattering geometry that is implemented in \texttt{NuMagSANS}. The neutron optical elements (polarizer, spin flipper, analyzer) that are required to measure the spin-resolved SANS cross sections are not drawn. The external magnetic field $\mathbf{B}_0 \parallel \mathbf{e}_z$ is here applied perpendicular to the wave vector $\mathbf{k}_0 \parallel \mathbf{e}_x$ of the incident neutron beam ($\mathbf{B}_0 \perp \mathbf{k}_0$). The momentum-transfer or scattering vector $\mathbf{q}$ is defined as the difference between $\mathbf{k}_0$ and $\mathbf{k}_1$, i.e., $\mathbf{q} = \mathbf{k}_0 - \mathbf{k}_1$. Its magnitude for elastic scattering, $q = \frac{4\pi}{\lambda} \sin\psi$, depends on the average neutron wavelength $\lambda$ and on the scattering angle $2\psi$. The angle $\theta = \angle(\mathbf{q}, \mathbf{B}_0)$ is used to describe the angular anisotropy of the scattering pattern on the two-dimensional detector. In small-angle approximation the component of $\mathbf{q}$ along $\mathbf{k}_0$ is small compared to the other two components, so that $\mathbf{q} \cong \{ 0, q_y, q_z \}$.}
\label{fig1}
\end{figure}

In this section, we display the basic and well-known expressions for the SANS observables, which include the various unpolarized, polarized, chiral, and nuclear-magnetic interference cross sections as well as the ensuing correlation and pair-distance distribution functions. We emphasize that nuclear spin-dependent scattering is not taken into account in \texttt{NuMagSANS}. Fig.~\ref{fig1} depicts the scattering geometry that is implemented in \texttt{NuMagSANS}: the incident neutron beam propagates along the \(x\)~axis, with the two-dimensional detector placed in the \(y\)-\(z\)-plane (see inset in Fig.~\ref{fig1}). This implies that, in small-angle  approximation, the scattering vector $\mathbf{q}$ has only components in the \(y\)-\(z\)-plane. The relevant SANS expressions for the two most-often employed geometries, where the applied magnetic field $\mathbf{B}_0$ is perpendicular ($\mathbf{B}_0 \perp \mathbf{k}_0$) or parallel ($\mathbf{B}_0 \parallel \mathbf{k}_0$) to the incident neutron beam are explicitly shown below. The theoretical framework for magnetic neutron scattering is thoroughly discussed in \cite{lovesey1984theory}, \cite{maleyev2002}, \cite{squires2012introduction}, \cite{boothroyd2016neutron}, and \cite{michelsbook}.

\subsection{Basic SANS cross sections and conventions}

We begin by introducing the continuum Fourier transforms of the (scalar) nuclear scattering-length density \(N(\mathbf{r}) = N(x,y,z)\) and the magnetization vector field \(\mathbf{M}(\mathbf{r}) = \{ M_x(\mathbf{r}), M_y(\mathbf{r}), M_z(\mathbf{r}) \}\):
\begin{align}
	\widetilde{N}(\mathbf{q}) &= \frac{1}{(2\pi)^{3/2}} \int_{\mathbb{R}^3} N(\mathbf{r}) \exp(-i \mathbf{q}\cdot\mathbf{r}) \, d^3r , \\
    \widetilde{\mathbf{M}}(\mathbf{q}) &= \frac{1}{(2\pi)^{3/2}} \int_{\mathbb{R}^3} \mathbf{M}(\mathbf{r}) \exp(-i \mathbf{q}\cdot\mathbf{r}) \, d^3r .
\end{align}
These Fourier transforms are functions of the wave vector $\mathbf{q} = \{ q_x, q_y, q_z \}$, whereas the magnetic SANS cross section (see below) is a function of the scattering vector, say $\mathbf{q}'$, which is the difference between the incident ($\mathbf{k}_0$) and scattered ($\mathbf{k}_1$) wave vectors of the neutron (compare Fig.~\ref{fig1}). In contrast to $\mathbf{q}$, the quantity $\mathbf{q}'$ is not a wave vector, since its magnitude for elastic scattering is not given by $2\pi/\lambda$. However, the SANS cross section at the scattering vector $\mathbf{q}'$ depends exclusively on the Fourier components of the magnetization at the wave vector $\mathbf{q} = \mathbf{q}'$, and a consistent separate handling of the symbols $\mathbf{q}$ and $\mathbf{q}'$ would unnecessarily encumber the discussion. In agreement with common usage in the neutron scattering literature, we ignore the distinction between the two quantities and use the symbol $\mathbf{q}$ to denote both the wave vector and the scattering vector. In \texttt{NuMagSANS} the above Fourier integrals are evaluated by discrete sums (see Section~\ref{sec:SummationApproach}).

The nuclear and (unpolarized) magnetic scattering contributions are then expressed as~\cite{squires2012introduction}:
\begin{align}
\frac{d\Sigma_{\mathrm{N}}}{d\Omega} &= \frac{8\pi^3}{V} \left| \widetilde{N} \right|^2 = \frac{8\pi^3}{V} \left( \widetilde{N} \cdot \widetilde{N}^\ast \right) ,
\label{eq:NuclearSANS}
\\
\frac{d\Sigma_{\mathrm{M}}}{d\Omega} &= \frac{8\pi^3}{V} b_{\mathrm{H}}^2 \left| \widetilde{\mathbf{Q}} \right|^2 = \frac{8\pi^3}{V} b_{\mathrm{H}}^2 \left( \widetilde{\mathbf{Q}} \cdot \widetilde{\mathbf{Q}}^\ast \right) = \frac{8\pi^3}{V} b_{\mathrm{H}}^2 \left( \left| \widetilde{Q}_x \right|^2 + \left| \widetilde{Q}_y \right|^2 + \left| \widetilde{Q}_z \right|^2 \right),
\label{eq:MagneticSANS}
\end{align}
where $V$ is the scattering volume, \(b_{\mathrm{H}} = 2.91 \times 10^{8} \, \mathrm{A}^{-1} \mathrm{m}^{-1}\) denotes the atomic magnetic scattering length in the small-angle regime, `$^\ast$' refers to the complex conjugate, and
\begin{align}
\widetilde{\mathbf{Q}}(\mathbf{q}) = \hat{\mathbf{q}} \times \left( \hat{\mathbf{q}} \times \widetilde{\mathbf{M}}(\mathbf{q}) \right)
\label{eq:HJdef}
\end{align}
is known as the Halpern-Johnson or magnetic interaction vector. In \texttt{NuMagSANS}, the incident neutron wave vector $\mathbf{k}_0$ is fixed to be parallel to the $x$~direction of a Cartesian laboratory coordinate system and the detector plane is spanned by the $y$-$z$~directions (compare Fig.~\ref{fig1}). This implies that, in the small-angle approximation, the unit scattering vector $\hat{\mathbf{q}}$ takes on the following form:
\begin{align}
\hat{\mathbf{q}} = \frac{\mathbf{q}}{|\mathbf{q}|} = \{ 0, \sin\theta, \cos\theta \} ,
\label{qdef}
\end{align}
which results in:
\begin{align}
\widetilde{\mathbf{Q}} = \left\{ \begin{array}{c} 
\widetilde{Q}_x \\ 
\widetilde{Q}_y \\ 
\widetilde{Q}_z 
\end{array} \right\} 
= \left\{ \begin{array}{c} 
- \widetilde{M}_x \\ 
- \widetilde{M}_y \cos^2\theta + \widetilde{M}_z \sin\theta \cos\theta \\ 
  \widetilde{M}_y \sin\theta \cos\theta - \widetilde{M}_z \sin^2\theta \end{array} \right\} .
\label{eq:HJdef1}
\end{align}

One of the first inputs the user must specify in \texttt{NuMagSANS} is the direction of the unit vector $\hat{\mathbf{P}}$. This quantity determines the orientation of the applied magnetic (guide) field $\mathbf{B}_0$ with respect to the wave vector $\mathbf{k}_0$ of the incoming beam. For example, $\hat{\mathbf{P}} = \{ 0, 0, 1 \}$ for $\mathbf{B}_0 \parallel \mathbf{e}_z \perp \mathbf{k}_0$, while $\hat{\mathbf{P}} = \{ 1, 0, 0 \}$ for $\mathbf{B}_0 \parallel \mathbf{e}_x \parallel \mathbf{k}_0$. With this in mind, and for the purpose of computational efficiency, we introduce the following quantity
\begin{align}
\frac{d\Sigma_{\mathrm{P}}}{d\Omega} &= \frac{8\pi^3}{V} b_{\mathrm{H}}^2 \left| \hat{\mathbf{P}} \cdot \widetilde{\mathbf{Q}} \right|^2 ,
\label{eq:PSANSdef}
\end{align}
which represents the projection of the Halpern-Johnson vector $\widetilde{\mathbf{Q}}$ onto the polarization unit vector $\hat{\mathbf{P}}$. In particular, $d\Sigma_{\mathrm{P}} / d\Omega$ is an auxiliary quantity that is not directly accessible in a SANS experiment and is typically not analyzed separately. However, it serves as a fundamental building block for the additive construction of directly measurable SANS cross sections, thereby enabling a higher degree of modularity and reusability in theoretical formulations and numerical implementations (see below). For a transverse scattering geometry $\mathbf{B}_0 \parallel \mathbf{e}_z \perp \mathbf{k}_0$, we have $d\Sigma_{\mathrm{P}} / d\Omega \propto | \widetilde{Q}_z|^2$, whereas for the longitudinal geometry $\mathbf{B}_0 \parallel \mathbf{e}_x \parallel \mathbf{k}_0$, it follows that $d\Sigma_{\mathrm{P}} / d\Omega \propto | \widetilde{Q}_x|^2$. More specifically,

\paragraph{$\mathbf{B}_0 \parallel \mathbf{e}_z \perp \mathbf{k}_0$}

\begin{align}
    \frac{d\Sigma_{\mathrm{P}}}{d\Omega} = \frac{8\pi^3}{V} b_{\mathrm{H}}^2 \left( |\widetilde{M}_y|^2 \sin^2\theta \cos^2\theta + |\widetilde{M}_z|^2 \sin^4\theta - (\widetilde{M}_y\widetilde{M}_z^{\ast} + \widetilde{M}_y^{\ast} \widetilde{M}_z) \sin^3\theta \cos\theta \right) ,
\end{align}

\paragraph{$\mathbf{B}_0 \parallel \mathbf{e}_x \parallel \mathbf{k}_0$}

\begin{align}
    \frac{d\Sigma_{\mathrm{P}}}{d\Omega} = \frac{8\pi^3}{V} b_{\mathrm{H}}^2 \; |\widetilde{M}_x|^2 .
\end{align}

\subsection{POLARIS cross sections}

In uniaxial or longitudinal polarization analysis experiments (denoted as POLARIS in the SANS community), one has access to four neutron scattering cross sections (spin channels). These include non-spin-flip and spin-flip processes. POLARIS experiments employ neutron optical elements to polarize the beam (relative to the magnetic guide field) before the sample, to flip the neutron spin direction before the sample, and to determine its polarization after the scattering process. The four partial cross sections are denoted as $d\Sigma^{++} / d\Omega$, $d\Sigma^{--} / d\Omega$, $d\Sigma^{+-} / d\Omega$, and $d\Sigma^{-+} / d\Omega$, where the first superscript to $d\Sigma / d\Omega$ (e.g., $+$) refers to the spin state of the incident neutrons, while the second (e.g., $-$) specifies the spin state of the scattered neutrons. The direction of $\hat{\mathbf{P}} \parallel \mathbf{B}_0$ specifies the quantization axis for neutron spins.

The \(+-\) and \(-+\) spin-flip cross sections can be expressed as follows:
\begin{align}
\frac{d\Sigma^{+-}}{d\Omega} &= \frac{d\Sigma_{\mathrm{sf}}}{d\Omega} + \frac{d\Sigma_{\chi}}{d\Omega} , \\
\frac{d\Sigma^{-+}}{d\Omega} &= \frac{d\Sigma_{\mathrm{sf}}}{d\Omega} - \frac{d\Sigma_{\chi}}{d\Omega} ,
\end{align}
where 
\begin{equation}
\frac{d\Sigma_{\mathrm{sf}}}{d\Omega} = \frac{d\Sigma_{\mathrm{M}}}{d\Omega} - \frac{d\Sigma_{\mathrm{P}}}{d\Omega}
\label{eq:sfdef}
\end{equation}
represents the {\it polarization-independent} part of the spin-flip SANS cross section, and
\begin{align}
\frac{d\Sigma_{\chi}}{d\Omega} = -i \frac{8\pi^3}{V} b_{\mathrm{H}}^2 \, \hat{\mathbf{P}} \cdot \left( \widetilde{\mathbf{Q}} \times \widetilde{\mathbf{Q}}^\ast \right) = -i \frac{8\pi^3}{V} b_{\mathrm{H}}^2 \, \chi,
\label{eq:ChiralSANS}
\end{align}
is the so-called chiral SANS cross section that depends on the polarization of the incident beam ($i^2 = -1$). Equation~(\ref{eq:sfdef}) shows that $d\Sigma_{\mathrm{sf}} / d\Omega \propto (| \widetilde{Q}_x|^2 + | \widetilde{Q}_y|^2)$ for $\mathbf{B}_0 \parallel \mathbf{e}_z \perp \mathbf{k}_0$, while $d\Sigma_{\mathrm{sf}} / d\Omega \propto (| \widetilde{Q}_y|^2 + | \widetilde{Q}_z|^2)$ for $\mathbf{B}_0 \parallel \mathbf{e}_x \parallel \mathbf{k}_0$. Equation~(\ref{eq:ChiralSANS}) also defines the function $\chi(\mathbf{q}) = \hat{\mathbf{P}} \cdot (\widetilde{\mathbf{Q}} \times \widetilde{\mathbf{Q}}^\ast)$, which for $\mathbf{B}_0 \parallel \mathbf{e}_z \perp \mathbf{k}_0$ equals
\begin{align}
\chi = \widetilde{Q}_x \widetilde{Q}_y^{\ast} - \widetilde{Q}_x^{\ast} \widetilde{Q}_y = \left( \widetilde{M}_x \widetilde{M}_y^{\ast} - \widetilde{M}_x^{\ast} \widetilde{M}_y \right) \cos^2\theta - \left( \widetilde{M}_x \widetilde{M}_z^{\ast} - \widetilde{M}_x^{\ast} \widetilde{M}_z \right) \sin\theta \cos\theta .
\end{align}
We note that $\chi(\mathbf{q})$ vanishes at complete magnetic saturation ($M_x = M_y = 0$), or for purely real-valued or purely imaginary magnetization Fourier components (irrespective of the value of the field)~\cite{Michels2016,sinaga2024neutron}. For example, for the case where the magnetization distribution is an even function of the position, i.e., $\mathbf{M}(\mathbf{r}) = \mathbf{M}(- \mathbf{r})$, the corresponding Fourier transform $\widetilde{\mathbf{M}}(\mathbf{q})$ is also an even and real-valued function~\cite{OranBrigham}, with the consequence that $\chi(\mathbf{q})$ vanishes. Similarly, if $\mathbf{M}(\mathbf{r})$ is an odd function, then $\widetilde{\mathbf{M}}(\mathbf{q})$ is an odd and imaginary function, and $\chi=0$. Note also that $\chi=0$ for $\mathbf{B}_0 \parallel \mathbf{e}_x \parallel \mathbf{k}_0$.

The \(++\) and \(--\) SANS cross sections are computed in \texttt{NuMagSANS} as:
\begin{align}
\frac{d\Sigma^{++}}{d\Omega} &= \frac{d\Sigma_{\mathrm{N}}}{d\Omega} + \frac{d\Sigma_{\mathrm{NM}}}{d\Omega} + \frac{d\Sigma_{\mathrm{P}}}{d\Omega} , \\
\frac{d\Sigma^{--}}{d\Omega} &= \frac{d\Sigma_{\mathrm{N}}}{d\Omega} - \frac{d\Sigma_{\mathrm{NM}}}{d\Omega} + \frac{d\Sigma_{\mathrm{P}}}{d\Omega} ,
\end{align}
where $d\Sigma_{\mathrm{N}} / d\Omega$ is the nuclear coherent SANS, $d\Sigma_{\mathrm{P}} / d\Omega$ has been defined by equation~(\ref{eq:PSANSdef}), and
\begin{align}
\frac{d\Sigma_{\mathrm{NM}}}{d\Omega} = \frac{8\pi^3}{V} b_{\mathrm{H}} \, \hat{\mathbf{P}} \cdot \left( \widetilde{N} \widetilde{\mathbf{Q}}^{\ast} + \widetilde{N}^{\ast} \widetilde{\mathbf{Q}} \right)
\label{eq:NMdef}
\end{align}
denote the nuclear-magnetic interference terms. For $\hat{\mathbf{P}} \parallel \mathbf{B}_0 \parallel \mathbf{e}_z \perp \mathbf{k}_0$, we have $d\Sigma_{\mathrm{NM}} / d\Omega \propto \widetilde{N} \widetilde{Q}_z$, while $d\Sigma_{\mathrm{NM}} / d\Omega \propto \widetilde{N} \widetilde{Q}_x$ for $\hat{\mathbf{P}} \parallel \mathbf{B}_0 \parallel \mathbf{e}_x \parallel \mathbf{k}_0$.

In the following, we display the POLARIS SANS cross sections for a perpendicular $\mathbf{B}_0 \parallel \mathbf{e}_z \perp \mathbf{k}_0$ and the parallel $\mathbf{B}_0 \parallel \mathbf{e}_x \parallel \mathbf{k}_0$ scattering geometry.

\paragraph{$\mathbf{B}_0 \parallel \mathbf{e}_z \perp \mathbf{k}_0$}

\begin{eqnarray}
\label{nsfperp}
\frac{d \Sigma^{\pm \pm}}{d \Omega} = \frac{8 \pi^3}{V} b_{\mathrm{H}}^2 \left( b_{\mathrm{H}}^{-2} |\widetilde{N}|^2 + |\widetilde{M}_y|^2 \sin^2\theta \cos^2\theta + |\widetilde{M}_z|^2 \sin^4\theta \right. \nonumber \\ \left. - (\widetilde{M}_y \widetilde{M}_z^{\ast} + \widetilde{M}_y^{\ast} \widetilde{M}_z) \sin^3\theta \cos\theta \right. \nonumber \\ \left. \mp b_{\mathrm{H}}^{-1} (\widetilde{N} \widetilde{M}_z^{\ast} + \widetilde{N}^{\ast} \widetilde{M}_z) \sin^2\theta \pm b_{\mathrm{H}}^{-1} (\widetilde{N} \widetilde{M}_y^{\ast} + \widetilde{N}^{\ast} \widetilde{M}_y) \sin\theta \cos\theta \right) ,
\end{eqnarray}

\begin{eqnarray}
\label{sfperp}
\frac{d \Sigma^{\pm \mp}}{d \Omega} = \frac{8 \pi^3}{V} b_{\mathrm{H}}^2 \left( |\widetilde{M}_x|^2 + |\widetilde{M}_y|^2 \cos^4\theta + |\widetilde{M}_z|^2 \sin^2\theta \cos^2\theta \nonumber \right. \\ \left. - (\widetilde{M}_y \widetilde{M}_z^{\ast} + \widetilde{M}_y^{\ast} \widetilde{M}_z) \sin\theta \cos^3\theta \mp i \chi \right) .
\end{eqnarray}
Although the spin-flip SANS cross section does not depend on the longitudinal Fourier component of the Halpern-Johnson vector ($\widetilde{Q}_z$ in the $\hat{\mathbf{P}} \parallel \mathbf{B}_0 \parallel \mathbf{e}_z \perp \mathbf{k}_0$ geometry), equation~(\ref{sfperp}) demonstrates that $d \Sigma^{\pm \mp} / d \Omega$ does depend on the longitudinal Fourier component $\widetilde{M}_z$ of the magnetization. This term gives rise to a pronounced $\sin^2\theta \cos^2\theta$ angular anisotropy of the spin-flip SANS cross section in the saturated state ($M_x = M_y = 0$).

\paragraph{$\mathbf{B}_0 \parallel \mathbf{e}_x \parallel \mathbf{k}_0$}

\begin{eqnarray}
\label{nsfpara}
\frac{d \Sigma^{\pm \pm}}{d \Omega} = \frac{8 \pi^3}{V} b_{\mathrm{H}}^2 \left( b_{\mathrm{H}}^{-2} |\widetilde{N}|^2 + |\widetilde{M}_x|^2 \mp b_{\mathrm{H}}^{-1} (\widetilde{N} \widetilde{M}_x^{\ast} + \widetilde{N}^{\ast} \widetilde{M}_x) \right) ,
\end{eqnarray}

\begin{eqnarray}
\label{sfpara}
\frac{d \Sigma^{\pm \mp}}{d \Omega} = \frac{8 \pi^3}{V} b_{\mathrm{H}}^2 \left( |\widetilde{M}_y|^2 \cos^2\theta + |\widetilde{M}_z|^2 \sin^2\theta - (\widetilde{M}_y \widetilde{M}_z^{\ast} + \widetilde{M}_y^{\ast} \widetilde{M}_z) \sin\theta \cos\theta \right) .
\end{eqnarray}
Due to the neglect of nuclear spin-dependent SANS, and since $\chi(\mathbf{q}) = 0$ for $\mathbf{B}_0 \parallel \mathbf{e}_x \parallel \mathbf{k}_0$, $d \Sigma^{\pm \mp} / d \Omega$ is independent of the incoming polarization for $\mathbf{B}_0 \parallel \mathbf{e}_x \parallel \mathbf{k}_0$.

We emphasize that for the determination of the POLARIS cross sections (equations~(\ref{nsfperp})$-$(\ref{sfpara})), it is generally necessary to measure the four partial cross sections to correct for spin leakage between the different channels~\cite{wildes06}. Such corrections can, e.g., be achieved by means of the Pol-Corr~\cite{krycka2012a,krycka2012b} and GRASP~\cite{dewhurst2023} software tools.

\subsection{SANSPOL cross sections}

The half-polarized SANS cross sections (so-called SANSPOL) are expressed in \texttt{NuMagSANS} as:
\begin{align}
\frac{d\Sigma^{+}}{d\Omega} &= \frac{d\Sigma^{++}}{d\Omega} + \frac{d\Sigma^{+-}}{d\Omega}, \\
\frac{d\Sigma^{-}}{d\Omega} &= \frac{d\Sigma^{--}}{d\Omega} + \frac{d\Sigma^{-+}}{d\Omega}.
\end{align}
These cross sections combine non-spin-flip and spin-flip contributions. Using equations~(\ref{nsfperp})$-$(\ref{sfpara}), we obtain for the perpendicular ($\mathbf{B}_0 \parallel \mathbf{e}_z \perp \mathbf{k}_0$) and parallel ($\mathbf{B}_0 \parallel \mathbf{e}_x \parallel \mathbf{k}_0$) scattering geometries:

\paragraph{$\mathbf{B}_0 \parallel \mathbf{e}_z \perp \mathbf{k}_0$}

\begin{eqnarray}
\label{sanspolperp}
\frac{d \Sigma^{\pm}}{d \Omega} = \frac{8 \pi^3}{V} b_{\mathrm{H}}^2 \left( b_{\mathrm{H}}^{-2} |\widetilde{N}|^2 + |\widetilde{M}_x|^2 + |\widetilde{M}_y|^2 \cos^2\theta + |\widetilde{M}_z|^2 \sin^2\theta \right. \nonumber \\ \left. - (\widetilde{M}_y \widetilde{M}_z^{\ast} + \widetilde{M}_y^{\ast} \widetilde{M}_z) \sin\theta \cos\theta \mp b_{\mathrm{H}}^{-1} (\widetilde{N} \widetilde{M}_z^{\ast} + \widetilde{N}^{\ast} \widetilde{M}_z) \sin^2\theta \right. \nonumber \\ \left. \pm b_{\mathrm{H}}^{-1} (\widetilde{N} \widetilde{M}_y^{\ast} + \widetilde{N}^{\ast} \widetilde{M}_y) \sin\theta \cos\theta \mp i \chi \right) ,
\end{eqnarray}

\paragraph{$\mathbf{B}_0 \parallel \mathbf{e}_x \parallel \mathbf{k}_0$}

\begin{eqnarray}
\label{sanspolpara}
\frac{d \Sigma^{\pm}}{d \Omega} = \frac{8 \pi^3}{V} b_{\mathrm{H}}^2 \left( b_{\mathrm{H}}^{-2} |\widetilde{N}|^2 + |\widetilde{M}_x|^2 + |\widetilde{M}_y|^2 \cos^2\theta + |\widetilde{M}_z|^2 \sin^2\theta \right. \nonumber \\ \left. - (\widetilde{M}_y \widetilde{M}_z^{\ast} + \widetilde{M}_y^{\ast} \widetilde{M}_z) \sin\theta \cos\theta \mp b_{\mathrm{H}}^{-1} (\widetilde{N} \widetilde{M}_x^{\ast} + \widetilde{N}^{\ast} \widetilde{M}_x) \right) . 
\end{eqnarray}

\subsection{Unpolarized SANS cross sections}
\label{unpolarized}

Finally, the unpolarized SANS cross section is computed according to:
\begin{align}
\frac{d\Sigma}{d\Omega} &= \frac{1}{2} \left( \frac{d\Sigma^{++}}{d\Omega} + \frac{d\Sigma^{--}}{d\Omega} + \frac{d\Sigma^{+-}}{d\Omega} + \frac{d\Sigma^{-+}}{d\Omega} \right) \nonumber \\
&= \frac{1}{2} \left( \frac{d\Sigma^{+}}{d\Omega} + \frac{d\Sigma^{-}}{d\Omega} \right) .
\end{align}

\paragraph{$\mathbf{B}_0 \parallel \mathbf{e}_z \perp \mathbf{k}_0$}

\begin{eqnarray}
\label{sigmasansperpunpol}
\frac{d \Sigma}{d \Omega} = \frac{8 \pi^3}{V} b_{\mathrm{H}}^2 \left( b_{\mathrm{H}}^{-2} |\widetilde{N}|^2 +  |\widetilde{M}_x|^2 + |\widetilde{M}_y|^2 \cos^2\theta + |\widetilde{M}_z|^2 \sin^2\theta \nonumber \right. \\ \left. - (\widetilde{M}_y \widetilde{M}_z^{\ast} + \widetilde{M}_y^{\ast} \widetilde{M}_z) \sin\theta \cos\theta \right) .
\end{eqnarray}

\paragraph{$\mathbf{B}_0 \parallel \mathbf{e}_x \parallel \mathbf{k}_0$}

\begin{eqnarray}
\label{sigmasansparaunpol}
\frac{d \Sigma}{d \Omega} = \frac{8 \pi^3}{V} b_{\mathrm{H}}^2 \left( b_{\mathrm{H}}^{-2} |\widetilde{N}|^2 + |\widetilde{M}_x|^2 + |\widetilde{M}_y|^2 \cos^2\theta + |\widetilde{M}_z|^2 \sin^2\theta \nonumber \right. \\ \left. - (\widetilde{M}_y \widetilde{M}_z^{\ast} + \widetilde{M}_y^{\ast} \widetilde{M}_z) \sin\theta \cos\theta \right) .
\end{eqnarray}

\subsection{Azimuthally averaged SANS cross sections}

All the computed two-dimensional SANS data sets $f(q, \theta)$ can be azimuthally averaged according to:
\begin{equation}
I(q) = \frac{1}{2\pi} \int_0^{2\pi} f(q, \theta) \, d\theta ,
\label{eq:iqazi}
\end{equation}
where \(\theta\) is the azimuthal angle on the detector plane (compare Fig.~\ref{fig1}). We note that the chiral SANS cross section $d\Sigma_{\chi} / d\Omega$, which exhibits a left-right asymmetry on the two-dimensional detector, always averages to zero using the integral~\eqref{eq:iqazi}. To analyze angular averages of $d\Sigma_{\chi} / d\Omega$ we suggest using the modal decomposition~\eqref{eq:CosineSinceIntensities} (see below). The integrals in equations~(\ref{eq:iqazi}), (\ref{eq:CosineSinceIntensities}), (\ref{eq:crrad}), and (\ref{eq:crav2d}) (see below) are numerically solved using the trapezoidal rule.

Beyond the azimuthal average, typically employed for the analysis of experimental data, \texttt{NuMagSANS} includes an advanced angular Fourier decomposition that enables a more detailed quantification of anisotropy in two-dimensional scattering patterns. 
More specifically, the two-dimensional scattering cross sections are decomposed into sine and cosine harmonics as
\begin{align}
    I_k^{\mathrm{s,c}}(q) &= \frac{2 - \delta_{k,0}}{2\pi} 
    \int_0^{2\pi} 
    f(q, \theta)\, \left\{\begin{matrix}\sin(k\theta)\\\cos(k\theta) \end{matrix}\right\}\, d\theta.
    \label{eq:CosineSinceIntensities}
\end{align}
These so-called modal intensities \( I_k^{\mathrm{c}}(q) \) and \( I_k^{\mathrm{s}}(q) \) represent the angular spectral components of the scattering cross section. Using the $q$-dependent intensities, \texttt{NuMagSANS} computes the related amplitudes
\begin{align}
    a_{k}^{\mathrm{s,c}} &= \sum_{i} I_{k}^{\mathrm{s,c}}(q_i) ,
\end{align}
with the following normalization
\begin{align}
    A_{k}^{\mathrm{s,c}} = \frac{a_{k}^{\mathrm{s,c}}}{\sum_{j=0}^{k_{\mathrm{max}}} \sqrt{|a_{j}^{\mathrm{s}}|^2 + |a_{j}^{\mathrm{c}}|^2}} .
    \label{eq:ModalAmplitudes}
\end{align}
These amplitudes allow for a direct anisotropy identification. In \texttt{NuMagSANS} the user can manually select a $k_{\mathrm{max}}$, which is the highest modal order up to which the SANS cross section is being decomposed ($0 \le k \le k_{\mathrm{max}}$). For example, a two-fold $\sin^2\theta$-anisotropy yields $A_{0}^{\mathrm{c}}/A_{2}^{\mathrm{c}} = - 1$, whereas for the case of a four-fold $\sin^2\theta \cos^2\theta$-anisotropy $A_{0}^{\mathrm{c}}/A_{4}^{\mathrm{c}} = - 1$.

\subsection{Correlation and pair-distance distribution functions}

Azimuthally averaged SANS cross sections \( I(q) \) can be further analyzed by applying the inverse Fourier transform to obtain real-space correlation functions and pair-distance distribution functions. These quantities provide insights into the internal structure and characteristic length scales of the sample. The correlation function \( c(r) \) and the pair-distance distribution function \( p(r) \) are calculated by numerically solving the following equations:
\begin{align}
c(r) &= \frac{1}{2\pi^2} \int_0^\infty I(q)\, j_0(q r)\, q^2\, dq, 
\label{eq:crrad} \\
p(r) &= r^2\, c(r),
\label{eq:prrad}
\end{align}
where \( j_0(x) = \sin(x)/x \) denotes the zeroth-order spherical Bessel function. 
For detailed discussions of the properties of \( p(r) \) and for practical aspects of its computation via indirect Fourier transformation, we refer the reader to the review articles by \cite{glatter} and \cite{svergun03}.

In addition to the one-dimensional correlation functions, \texttt{NuMagSANS} also supports the computation of two-dimensional correlation functions, which are defined as the inverse two-dimensional Fourier transforms of the two-dimensional SANS cross sections \( \frac{d\Sigma}{d\Omega} = \frac{d\Sigma}{d\Omega}(q,\theta) \):
\begin{align}
    C(\rho,\alpha) = \int_{0}^{2\pi} \!\int_{0}^{\infty} 
    \frac{d\Sigma}{d\Omega}(q,\theta)\,
    \exp\left(i q \rho \cos(\theta - \alpha)\right)\, q\, dq\, d\theta .
    \label{eq:crav2d}
\end{align}
We emphasize that the one-dimensional correlation function \( c(r) \) and the azimuthal average of the two-dimensional correlation function \( C(\rho,\alpha) \) (with respect to $\alpha$) are connected via an Abel transform. Therefore, \( c(r) \) is not just simply the azimuthal average of \( C(\rho,\alpha) \)~\cite{michelsbook}.

\section{Program description: implementation and workflow}
\label{sec:ProgramOrgranization}

\begin{figure}[htb!]
\centering
\resizebox{0.80\columnwidth}{!}{\includegraphics{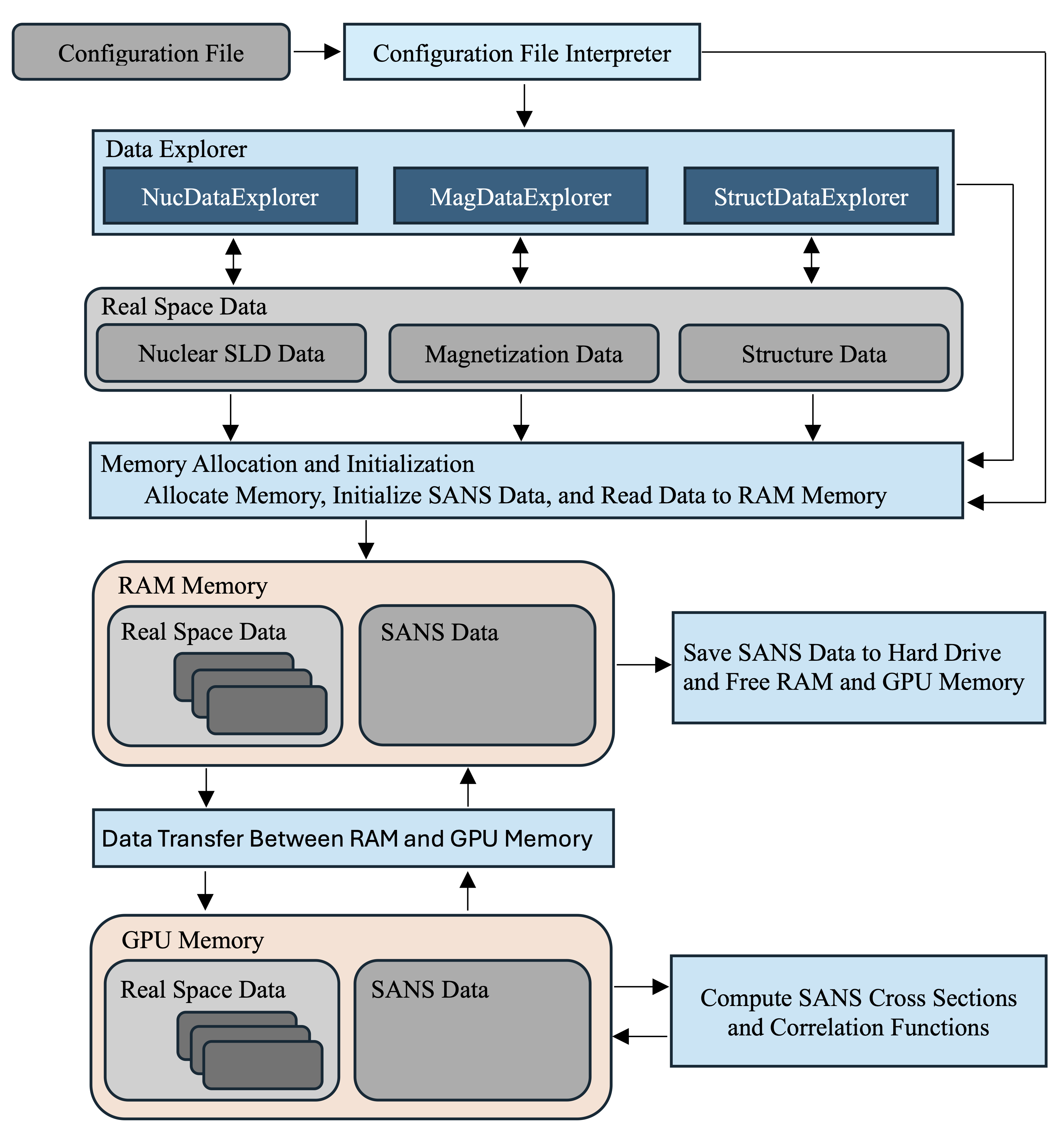}}
\caption{Schematic of the basic workflow of \texttt{NuMagSANS} (see the main text for explanations).}
\label{fig2}
\end{figure}

The implementation and workflow of \texttt{NuMagSANS} are organized into several modular components, as outlined in Fig.~\ref{fig2}. These components allow for a streamlined and efficient calculation of SANS observables based on the real-space nuclear scattering-length density, the magnetization distribution, and the structural (object position) data provided by the user. This section provides an overview of the overall workflow, followed by detailed discussions of the configuration file, real-space data organization, the summation methodology, and the SANS data computation.

As shown in Fig.~\ref{fig2}, the workflow begins with the \texttt{Configuration File}, which serves as the central input to specify the parameters and settings required for the SANS computations. The configuration file is parsed by the \texttt{Configuration File Interpreter}, which initializes the workflow by providing relevant parameters to other modules. 

Next, the \texttt{Data Explorer} modules (\texttt{NucDataExplorer}, \texttt{MagDataExplorer}, \texttt{StructDataExplorer}) manage the user-supplied nuclear scattering-length density, magnetization vector field, and structural data, collectively referred to as \texttt{Real Space Data}. This data are crucial for defining the physical system to be analyzed.

The real-space data are subsequently processed during the \texttt{Memory Allocation} and \texttt{Initialization} phase, where both the input real-space and the output SANS data structures are loaded into the RAM. Following this, the real-space data are transferred to the GPU memory to leverage CUDA for high-performance computation of the SANS observables. Once the calculations are complete, the resulting SANS data are saved to the hard drive, and both RAM and GPU memory are cleared to prepare for subsequent runs.

In the following subsections, we describe the role and structure of the configuration file (Section~\ref{subsec:ConfigurationFile}), the organization of the real-space input data (Section~\ref{subsec:RealSpaceData}), the summation scheme (Section~\ref{sec:SummationApproach}), and the organization of the output data for the SANS observables (Section~\ref{subsec:SANSData}).


\subsection{Configuration file}
\label{subsec:ConfigurationFile}

The configuration file serves as the central control unit for defining all input parameters, simulation settings, and data paths required for a \texttt{NuMagSANS} computation. It is a human-readable text file located in the main directory of the GitHub repository under the name \texttt{NuMagSANSInput.conf}. The file is parsed by the internal configuration interpreter, which initializes the workflow and allocates the corresponding memory structures. 

In this configuration file, the user specifies the following information:
\begin{itemize}
    \item the paths to the nuclear, magnetic, and structural (object position) data,
    \item the output directory name, where all simulation results are stored,
    \item whether nuclear, magnetic, or structural contributions are activated,
    \item the computational mode (atomistic or micromagnetic) to be applied,
    \item the saturation magnetization $M_{\mathrm{s}}$ (in A/m, micromagnetic simulation mode only),
    \item the proportionality factor $N_{\mathrm{s}}$ for the nuclear scattering-length density (in $\mathrm{m}^{-2}$, mesoscopic simulation mode only),
    \item the scaling factor $s$ for the lattice length scale; if the input data are in nm then $s=1$; if the input data are in meters then $s=10^{9}$; if the input data are in Angstrom then $s=10^{-1}$,
    \item the discretization volume $a_x, a_y, a_z$ used in the micromagnetic simulations (in nm),
    \item the total scattering volume $V$ (in $\mathrm{m}^3$), which is required for computing absolute SANS intensities,
    \item the maximum scattering vector range $q_\mathrm{max}$ (in $\mathrm{nm}^{-1}$); note that $q_\mathrm{min} = 0$ is fixed,
    \item the corresponding real-space range $r_\mathrm{max}$ (in nm) for the computation of correlation functions,
    \item the $q$, $\theta$, $r$, and $\alpha$-resolution (number of bins; a typical value is $1000$),
    \item two sample rotation angles, $\alpha$ and $\beta$ (in deg), describing rotations of the sample with respect to the laboratory coordinate system,
    \item and optional loop modes that enable automated batch runs for field-dependent, temperature-dependent, or time-dependent real-space data sets.
\end{itemize}

In the \textbf{atomistic simulation mode}, the parameters $a_x, a_y, a_z$ are irrelevant, as the atomic positions are treated as discrete point objects. Likewise, the saturation magnetization $M_{\mathrm{s}}$ is not used, since the atomic magnetic moments are expressed in units of the Bohr magneton $\mu_{\mathrm{B}} = 9.274 \times 10^{-24}~\mathrm{J/T}$, and the nuclear scattering length is given in units of $10^{-15}~\mathrm{m}$.

In the \textbf{micromagnetic simulation mode}, where the data typically originate from \texttt{MuMax3}-type simulations, the user must specify the simulation cell-size parameters $a_x, a_y, a_z$ (in nm) and the saturation magnetization $M_{\mathrm{s}}$ (in A/m), which serves as a scaling factor for the magnetization vectors. Similarly, for nuclear scattering-length density data in the mesoscopic mode, a proportionality factor $N_{\mathrm{s}}$ can be defined to match the physical scattering-length density contrast.

In the micromagnetic (mesoscopic) simulation mode, where an idealized continuous (nonperiodic) nuclear scattering length density and magnetization vector field are considered, the resulting SANS cross section does not exhibit Bragg reflections originating from the atomic crystal lattice. In contrast, as \texttt{NuMagSANS} handles numerical data sets (e.g., from micromagnetic solvers such as \texttt{MuMax3}), spurious Bragg peaks can appear due to the underlying numerical grid used in the simulations. To avoid these artifacts, the maximum accessible scattering vector $q_{\mathrm{max}}$ should be chosen so that these discretization-induced peaks are excluded. For instance, if the discretization is defined by $a_x = a_y = a_z = a$, the first artificial Bragg peak arising from the discrete real-space lattice (numerical grid) is expected to appear at
\[
q_{\mathrm{Bragg}} = \frac{2\pi}{a}.
\]
A suitable choice is to limit the data to the Nyquist condition:
\[
q_{\mathrm{max}} \simeq q_{\mathrm{Nyquist}} = \frac{\pi}{a}.
\]
This ensures that the SANS intensity is sampled within the physically significant Fourier range of the input field.

For the correlation functions, the maximum correlation distance $r_{\mathrm{max}}$ should be chosen to be equal to or larger than the largest geometrical distance represented in the input data. For correlation distances that exceed the largest geometrical distance in the user input data, the one-dimensional correlation functions and pair-distance distribution functions are expected to be zero. This condition can be used as a numerical consistency check within \texttt{NuMagSANS}. It should be emphasized that for the two-dimensional magnetic correlation functions this condition is in general not fulfilled, since the two-dimensional SANS cross section depends on the Fourier transform of the magnetic induction field $\mathbf{B}(\mathbf{r})$ of the sample, which can give rise to correlations both inside and outside of the sample (due to stray-field contributions).

In the case of polarized SANS, the direction of the polarization unit vector \(\hat{\mathbf{P}} = \{ P_x, P_y, P_z \}\) must be selected. Based on the choice of \(\hat{\mathbf{P}} \parallel \mathbf{B}_0\), the user can define the standard (most-often employed) scattering geometries: \(\hat{\mathbf{P}} = \{ 0, 0, 1 \}\) or \(\hat{\mathbf{P}} = \{ 0, 1, 0 \}\) (perpendicular) or \(\hat{\mathbf{P}} = \{1, 0, 0 \}\) (parallel), referring to the relative orientation of the polarization vector and the neutron beam.

The configuration file also allows for the definition of two sample rotation angles, $\alpha$ and $\beta$, which describe the rotations of the sample with respect to the laboratory coordinate system. The rotation matrix is defined as a composition of $z$- and $y$-axis rotations:
\begin{align}
    \mathbf{R}(\alpha,\beta) = \mathbf{R}_z(\beta)\, \mathbf{R}_y(\alpha)
    =
    \begin{Bmatrix}
        R_{11} & R_{12} & R_{13} \\
        R_{21} & R_{22} & R_{23} \\
        R_{31} & R_{32} & R_{33}
    \end{Bmatrix}.
\end{align}

A typical example is the micromagnetic simulation of a nanoparticle under an external magnetic field $\mathbf{B}_0$ applied along the $z$~direction. For this configuration, $\alpha = \beta = 0$ corresponds to the perpendicular scattering geometry, since the neutron beam is always assumed to be parallel to the $x$~axis, and the polarization vector \(\hat{\mathbf{P}}= \{0,0,1\}\) is directed along the applied field. To switch to the parallel scattering geometry ($\hat{\mathbf{P}} \parallel \mathbf{B}_0 \parallel \mathbf{k}_0$), the user sets $\alpha = 90^{\circ}$ and $\beta = 0$, which corresponds to a $90^{\circ}$ rotation around the $y$~axis, and the polarization direction must be set to \(\hat{\mathbf{P}}= \{1,0,0\}\).

\subsection{Real-space input data}
\label{subsec:RealSpaceData}

Next to the configuration file, \texttt{NuMagSANS} deals with three different types of input data: one for the nuclear structure, one for the magnetic structure, and one for the structural data describing the spatial arrangement of the objects (particles). The structure of the respective types of files is as follows:
\begin{itemize}
    \item nuclear scattering-length density data (NucData): $\{x^{kl}, \; y^{kl}, \; z^{kl}, \;\nu^{kl}\}$,
    \item magnetic data (MagData): $\{x^{kl}, \;  y^{kl}, \; z^{kl}, \; \mu_x^{kl}, \;\mu_y^{kl}, \;\mu_z^{kl}\}$,
    \item global object position data (StructData): $\{r_x^k, \; r_y^k, \; r_z^k\}$,
\end{itemize}
where the $x^{kl}, y^{kl}, z^{kl}$ specify the position data corresponding to the nuclear scattering-length data $\nu^{kl}$ or the magnetic vector data $\mu_x^{kl}, \mu_y^{kl}, \mu_z^{kl}$, respectively. The $r_x^{k}, r_y^{k}, r_z^{k}$ specify additional structural data for advanced analysis (see below). Possible file extensions are $.txt$, $.csv$, $.obj$, and the data files are always expected to have a columnar structure with a space as delimiter (see Table~\ref{tab:MagDataFormat}). In the configuration file, the user must select the types (nuclear, magnetic, structural) of input data (see Listing~\ref{lst:DataActivation}).

\begin{table}[h!]
\centering
\caption{Example of a magnetic data file (\texttt{MagData}) used by \texttt{NuMagSANS}. Each horizontal row lists the spatial coordinates $\{x,\; y, \; z\}$ of a voxel and the associated magnetization vector $\{m_x, \; m_y, \; m_z\}$. The example shown corresponds to the test data set of a uniformly magnetized nanoparticle, where all magnetization vectors point along the \(z\)~direction (see Section~\ref{sec:Example1}).}
\begin{tabular}{rrrrrr}
\hline
$x$ & $y$ & $z$ & $m_x$ & $m_y$ & $m_z$ \\
\hline
 $0.0$ & $-20.0$ & $0.0$ & $0.0$ &  $0.0$ & $1.0$ \\
$-6.0$ & $-19.0$ & $-1.0$ & $0.0$ & $0.0$ & $1.0$ \\
$-6.0$ & $-19.0$ & $0.0$ & $0.0$ & $0.0$ & $1.0$ \\
$-6.0$ & $-19.0$ & $1.0$ & $0.0$ & $0.0$ & $1.0$ \\
$-5.0$ & $-19.0$ & $-3.0$ & $0.0$ & $0.0$ & $1.0$ \\
$-5.0$ & $-19.0$ & $-2.0$ & $0.0$ & $0.0$ & $1.0$ \\
$-5.0$ & $-19.0$ & $-1.0$ & $0.0$ & $0.0$ & $1.0$ \\
$-5.0$ & $-19.0$ & $0.0$ & $0.0$ & $0.0$ & $1.0$ \\
$-5.0$ & $-19.0$ & $1.0$ & $0.0$ & $0.0$ & $1.0$ \\
$-5.0$ & $-19.0$ & $2.0$ & $0.0$ & $0.0$ & $1.0$ \\
$\vdots$  & $\vdots$ & $\vdots$ & $\vdots$ & $\vdots$ & $\vdots$ \\
\hline
\end{tabular}
\label{tab:MagDataFormat}
\end{table}

\begin{lstlisting}[caption={Example for the input data selection in the configuration file. Here, MagData is activated, and NucData and StructData are deactivated.}, label={lst:DataActivation}]
NucData_activate    = 0;   // activate nuclear data   (1 = on, 0 = off)
MagData_activate    = 1;   // activate magnetic data  (1 = on, 0 = off)
StructData_activate = 0;   // activate structure data (1 = on, 0 = off)
\end{lstlisting}

The input data indexing is organized as follows: 
\begin{itemize}
\item Object index (e.g., a particle with a certain orientation): $k = \{0, 1, 2, ..., K-1\}$,
\item Atom/cell index: $l = \{0, 1, 2, ...,  N_k -1 \}$,
\item Total number of objects (e.g., particles) in the sample: $K$,
\item Total number of atoms/cells per object: $N_k$,
\item Total number of atoms/cells per sample: $W = \sum_{k=0}^{K-1} N_k$,
\item Average number of atoms/cells per object: $\overline{N} = W/K$.
\end{itemize}

We reemphasize that in the atomistic simulation mode the atomic magnetic moment components $\mu_x^{kl}, \mu_y^{kl}, \mu_z^{kl}$ are expressed in units of the Bohr magneton $\mu_{\mathrm{B}} = 9.274 \times 10^{-24}$~J/T, whereas in the micromagnetic simulation mode the $\mu_x^{kl}, \mu_y^{kl}, \mu_z^{kl}$ are dimensionless; they are later normalized by the saturation magnetization $M_{\mathrm{s}}$ (in A/m), which is defined by the user in the configuration file. The cell magnetic moment is then computed as $\mu_{\mathrm{cell}} = M_{\mathrm{s}} V_{\mathrm{cell}}$, with the cell volume given by $V_{\mathrm{cell}}=a_x a_y a_z$. Likewise, the $\nu^{kl}$ come in units of $10^{-15} \, \mathrm{m}$ for atomistic simulations, and are dimensionless for the case of continuum/mesoscopic simulations; a scaling factor $N_{\rm s}$ (in units of m$^{-2}$, to be input in the configuration file) allows the user to express the $\nu^{kl}$ in units of m$^{-2}$. The corresponding nuclear cell scattering length is then given by $n_{\mathrm{cell}} = N_{\rm s} V_{\mathrm{cell}}$.

\begin{figure}[htb!]
\centering
\resizebox{0.70\columnwidth}{!}{\includegraphics{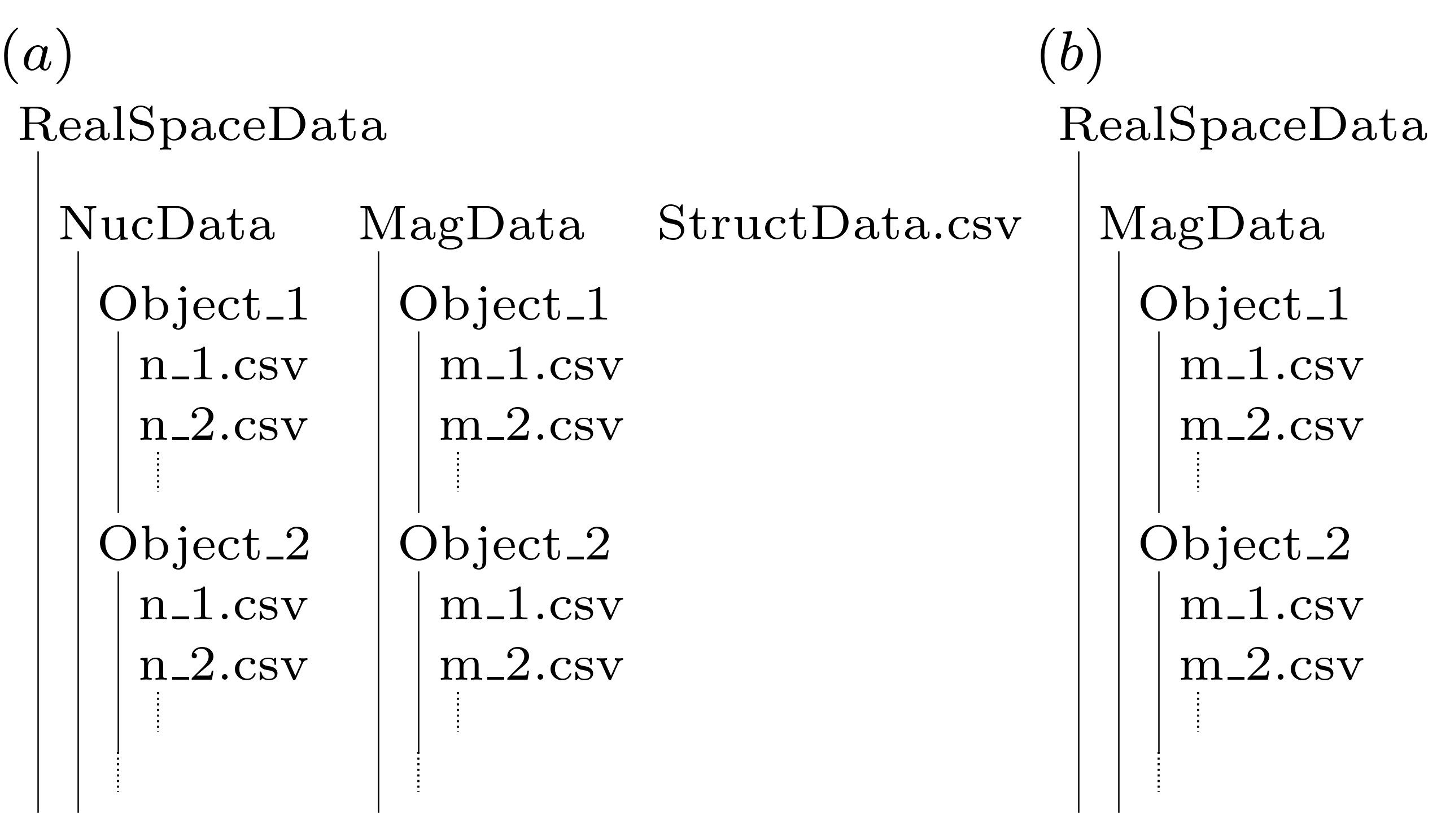}}
\caption{
Organization of the real-space input data used by \texttt{NuMagSANS}. ($a$)~General case in which all input channels are active: nuclear SLD data (\texttt{NucData}), magnetic vector-field data (\texttt{MagData}), and an optional structural description (\texttt{StructData.csv}) defining the spatial arrangement of multiple objects. This setup enables simulations ranging from fully dilute ensembles (no interparticle interference) to ordered or densely packed systems that exhibit structure-factor effects. Users may either represent many individual particles or encode the entire sample in a single object. ($b$)~Case with only magnetic data activated, corresponding to Listing~\ref{lst:DataActivation}. This situation corresponds to a dilute particle system, since no structural data is specified.}
\label{fig3}
\end{figure}

For the simulation of complex systems and big data, we need a well organized structure for the input data sets. In Fig.~\ref{fig3}, we show that in the main data directory we have two directories, \texttt{NucData} and \texttt{MagData}, and the \texttt{StructData}-file. A concrete example case is a dilute ensemble of 800 spherical iron nanoparticles with 1000 different applied magnetic fields $B_0$ along the hysteresis loop~\cite{adams2024framework}. For the 800 iron spheres~($=$~objects), each one having a different orientation of its magnetic anisotropy axis with respect to $\mathbf{B}_0$, this means that we have a \texttt{MagData}-directory that includes 800 directories (Object\_1 $\dots$ Object\_800), where each of these 800 directories stores 1000 magnetic structure data files (m\_1.csv $\dots$ m\_1000.csv) corresponding to the different applied fields. Note that the numbering of these directories and files must be continuous, starting from `1'.

The global object position data $r_x, r_y, r_z$ allow the user to perform an advanced analysis, where the effects between interacting and noninteracting objects can be studied. In the noninteracting case, the global object position data are not needed, and the \texttt{StructData.csv}-file can be excluded in the configuration file (see Listing~\ref{lst:DataActivation}). For the case of interacting objects, the user needs to generate its own global object position data that define the spatial arrangement of the objects. Examples for \texttt{StructData.csv}-files are provided in Section~\ref{sec:Example3} and on Zenodo~\cite{Adams2025_Example3}.

For the case of nonzero rotation angles $\alpha,\beta$ (see Section~\ref{subsec:ConfigurationFile}), the transformation of the global position data $x^{kl} + r_{x}^k, \; y^{kl} + r_{y}^k, \; z^{kl} + r_{z}^k$ is then performed as:
\begin{align}
    X^{kl} &= 
    R_{11} (x^{kl} + r_{x}^k) 
    + R_{12} (y^{kl} + r_{y}^k) 
    + R_{13} (z^{kl} + r_{z}^k) ,
    \\
    Y^{kl} &= 
    R_{21} (x^{kl} + r_{x}^k) 
    + R_{22} (y^{kl} + r_{y}^k) 
    + R_{23} (z^{kl} + r_{z}^k) ,
    \\
    Z^{kl} &= 
    R_{31} (x^{kl} + r_{x}^k) 
    + R_{32} (y^{kl} + r_{y}^k) 
    + R_{33} (z^{kl} + r_{z}^k).
\end{align}
Note that for a noninteracting (dilute) system, the $r_{x}^k, r_{y}^k, r_{z}^k$ are all equal to zero.

In \texttt{NuMagSANS} a fixed scattering geometry is implemented, where the incoming neutron beam is parallel to the $\mathbf{e}_x$~direction of a Cartesian laboratory coordinate system and the two-dimensional detector plane is spanned by the Fourier coordinates $q_y = q \sin\theta$ and $q_z = q \cos\theta$ (compare Fig.~\ref{fig1}). This means that \texttt{NuMagSANS} computes two-dimensional scattering cross sections as a function of the two polar Fourier coordinates $(q,\theta)$, which is a convenient choice for the further analysis (e.g., azimuthal averages). For the further analysis, it is convenient to define the following phase function $\Psi^{kl}(q,\theta)$
\begin{align}
    \Psi^{kl}(q,\theta) = q (Y^{kl} \sin(\theta) + Z^{kl} \cos\theta) ,
\end{align}
which is used in the Fourier summation approach.

\subsection{Summation approach}
\label{sec:SummationApproach}

The Fourier transform $\widetilde{\mathbf{M}}(\mathbf{q})$ of the magnetization vector field is expressed in units of $\mathrm{A m}^2$, whereas the Fourier transform $\widetilde{N}(\mathbf{q})$ of the nuclear scattering-length density is expressed in units of $\mathrm{m}$. The corresponding discrete-space Fourier transforms (DSFTs) are given by
\begin{align}
	\widetilde{N}(\mathbf{q}) &= \frac{1}{(2\pi)^{3/2}} \sum_{j} n_j  \exp(-i \mathbf{q}\cdot\mathbf{r}_j),\\
    \widetilde{\mathbf{M}}(\mathbf{q}) &= \frac{1}{(2\pi)^{3/2}} \sum_{j} \mathbf{m}_j \exp(-i \mathbf{q}\cdot\mathbf{r}_j),
\end{align}
where the index `$j$' enumerates the real-space discretization cells (or atoms) with coordinates $\mathbf{r}_j$. The vectors $\mathbf{m}_j$ represent the cell or atomic magnetic moments in units of $\mathrm{A\,m}^2$, and the $n_j$ are the corresponding nuclear scattering lengths in units of $\mathrm{m}$. This DSFT formulation serves as the basis for further developments in complex particle systems.

Since C++/CUDA does not support complex-number arithmetic in kernel code, the real and imaginary parts of the Fourier transforms are computed separately via discrete cosine and sine transforms according to:
\begin{align}
    \widetilde{\nu}^{R}(q,\theta) &= +\frac{1}{\sqrt{W \overline{N}}}\sum_{k=0}^{K-1}\sum_{l=0}^{N_k-1} \nu^{kl} \cos(\Psi^{kl}(q,\theta)) ,
    \\
     \widetilde{\nu}^{I}(q,\theta) &= -\frac{1}{\sqrt{W \overline{N}}}\sum_{k=0}^{K-1}\sum_{l=0}^{N_k-1} \nu^{kl} \sin(\Psi^{kl}(q,\theta)) ,
     \\
    \widetilde{\mu}_{\alpha}^{R}(q,\theta) &= +\frac{1}{\sqrt{W \overline{N}}}\sum_{k=0}^{K-1}\sum_{l=0}^{N_k-1} \mu_{\alpha}^{kl} \cos(\Psi^{kl}(q,\theta)) ,
    \\
    \widetilde{\mu}_{\alpha}^{I}(q,\theta) &= -\frac{1}{\sqrt{W \overline{N}}}\sum_{k=0}^{K-1}\sum_{l=0}^{N_k-1} \mu_{\alpha}^{kl} \sin(\Psi^{kl}(q,\theta)) ,
\end{align}
where $\alpha \in \{x, \; y, \; z\}$, and the Fourier sums are normalized by the factor $(W \overline{N})^{-1/2}$. These summations are executed in parallel on the GPU for each Fourier-space tuple $\{ q_i, \; \theta_j \}$, providing the primary performance gain compared to a sequential CPU implementation. The resulting nuclear ($\widetilde{\nu}$) and magnetic ($\widetilde{\mu}$) Fourier components are then used to construct the Halpern-Johnson vector and the corresponding unpolarized and polarized SANS cross sections.

To obtain the SANS cross section in absolute units of $\mathrm{cm}^{-1}$, the cross section data are rescaled prior to their export. For example, in the atomistic simulation mode, the scaling factor for the purely magnetic contribution reads
\begin{align}
\frac{d\Sigma_{\mathrm{M}}}{d\Omega} = W \overline{N}\, \frac{b_{\mathrm{H}}^2}{V} \mu_{\mathrm{B}}^2 \frac{dS_{\mathrm{M}}}{d\Omega},
\end{align}
where $\frac{dS_{\mathrm{M}}}{d\Omega}$ denotes the dimensionless, scaled magnetic SANS cross section.

We emphasize that the combination of the local position data $\mathbf{x}^{kl} = \{x^{kl}, \; y^{kl}, \; z^{kl}\}$ and the global object position data $\mathbf{r}^{k} = \{r_x^k, \; r_y^{k}, \; r_z^{k}\}$ generally leads to the following summation relation for the scattering amplitude (here given for the example of nuclear scattering):
\begin{align}
    \left|\sum_{k}\sum_{l} \nu^{kl} \exp(-i \mathbf{q}\cdot[\mathbf{x}^{kl} + \mathbf{r}^k])\right|^2 &= 
    \sum_{k}\left|\sum_{l} \nu^{kl} \exp(-i \mathbf{q}\cdot\mathbf{x}^{kl})\right|^2 \nonumber
    \\
    &+
     \sum_{k}\sum_{k'\neq k}\sum_{l}\sum_{l'} \nu^{kl} \nu^{k'l'} \exp(-i \mathbf{q}\cdot[(\mathbf{x}^{kl}+\mathbf{r}^k) - (\mathbf{x}^{k'l'}+\mathbf{r}^{k'}) ]).
\end{align}
The first term, with the summation over $k$ outside the absolute square brackets, describes the averaged intraparticle contributions, whereas the second term accounts for the interparticle interference between different objects $k\neq k'$.

By contrast, for a dilute or noninteracting system, where no \texttt{StructData} is provided by the user, the summation simplifies to~\cite{michelsbook}: 
\begin{align}
    \left|\sum_{k}\sum_{l} \nu^{kl} \exp(-i \mathbf{q}\cdot[\mathbf{x}^{kl} + \mathbf{r}^k])\right|^2\cong \sum_{k}\left|\sum_{l} \nu^{kl} \exp(-i \mathbf{q}\cdot\mathbf{x}^{kl})\right|^2 ,
\end{align}
where all interparticle interference terms are neglected.

\subsection{Fourier and real-space space output data}
\label{subsec:SANSData}

\begin{figure}[htb!]
\centering
\resizebox{0.60\columnwidth}{!}{\includegraphics{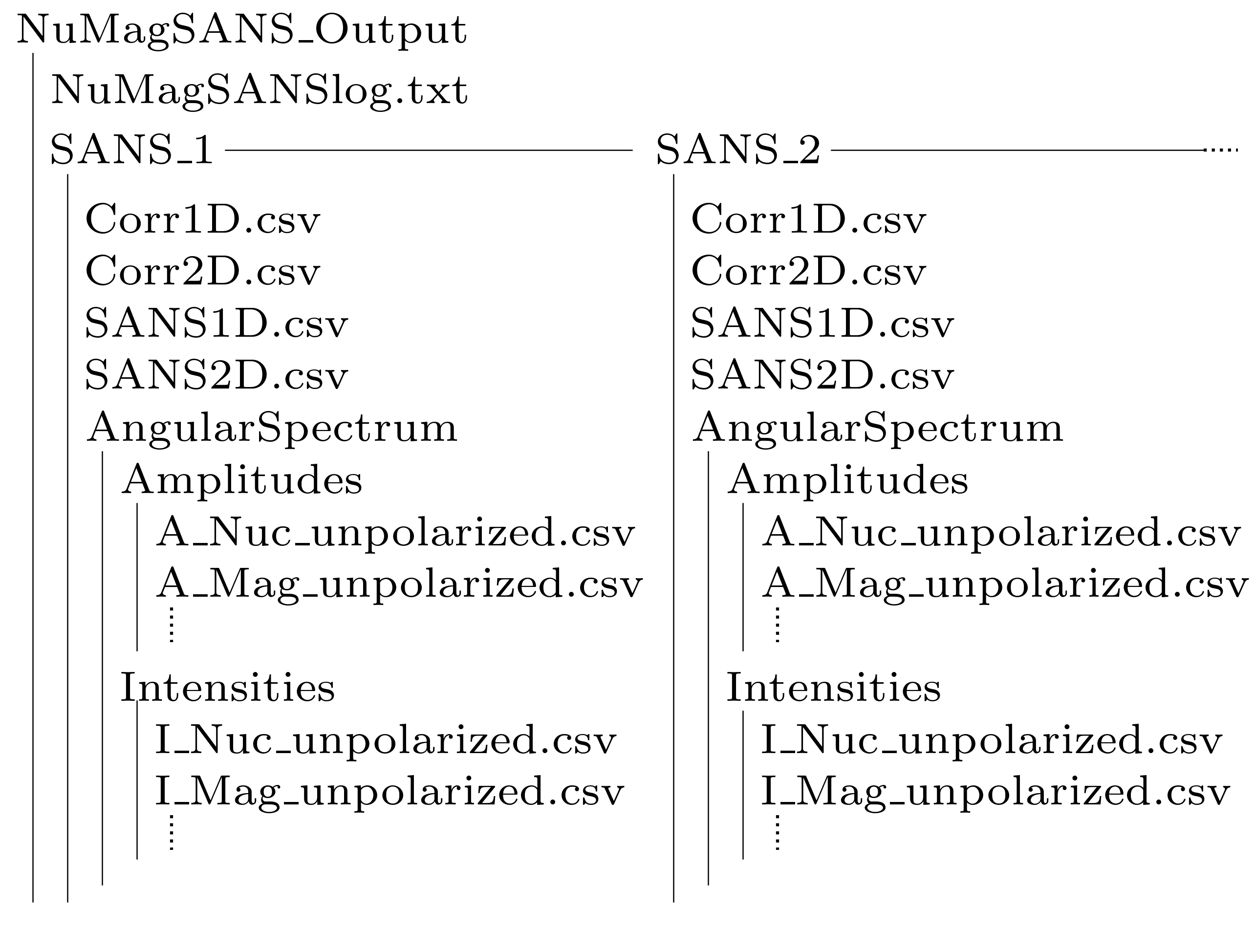}}
\caption{Directory structure of the \texttt{NuMagSANS\_Output} folder generated by \texttt{NuMagSANS}. The file \texttt{NuMagSANSlog.txt} acts as a metadata record of all program steps and the user-defined settings specified in \texttt{NuMagSANSInput.conf}. Depending on how many parameter sets were requested (e.g., applied field values), the output contains several subdirectories \texttt{SANS\_i}. Each \texttt{SANS\_i} directory includes the two-dimensional SANS cross section (\texttt{SANS2D.csv}); the azimuthally averaged intensity \(I(q)\) according to equation~\eqref{eq:iqazi} (\texttt{SANS1D.csv}); the two-dimensional real-space correlation function \(C(\rho,\alpha)\) from equation~\eqref{eq:crav2d} (\texttt{Corr2D.csv}); and the one-dimensional correlation and pair-distance distribution functions \(c(r)\) and \(p(r)\), defined in equations~\eqref{eq:crrad} and~\eqref{eq:prrad} (\texttt{Corr1D.csv}). Furthermore, each directory may contain an \texttt{AngularSpectrum} folder, which stores the modal sine and cosine intensities \(I_{k}^{\mathrm{s,c}}(q)\) computed from equation~\eqref{eq:CosineSinceIntensities}, and the corresponding normalized modal amplitudes \(A_{k}^{\mathrm{s,c}}\) defined in equation~\eqref{eq:ModalAmplitudes}.}
\label{fig4}
\end{figure}

The output data of \texttt{NuMagSANS} consist of four CSV files named \texttt{SANS1D.csv}, \texttt{SANS2D.csv}, \texttt{Corr1D.csv}, and \texttt{Corr2D.csv}, and an additional directory \texttt{AngularSpectrum} that contains the results computed by equations~\eqref{eq:CosineSinceIntensities} and \eqref{eq:ModalAmplitudes}. In the case of magnetic-field-dependent data, where the user data set comprises multiple real-space files (m\_1.csv, m\_2.csv, ...), and the loop mode is enabled, a separate output folder named SANS\_$i$ is created for each index `$i$' (e.g., magnetic field value), containing the corresponding set of output files (four CSV files), as illustrated in Fig.~\ref{fig4}.

The files \texttt{SANS1D.csv} and \texttt{SANS2D.csv} contain the azimuthally averaged and the two-dimensional magnetic SANS cross sections, respectively. The scattering vector magnitude $q$ is provided in units of nm$^{-1}$, the polar angle $\theta$ is in radians, and the scattering cross section is in units of cm$^{-1}$. Note that the \texttt{SANS2D.csv} file stores the two-dimensional SANS cross sections in a single-column format. For post-processing, the data must therefore be rearranged, for example using the \texttt{reshape} function in \textsc{Matlab} or the \texttt{reshape} function of the \texttt{NumPy} package in Python. The same logic applies to the two-dimensional correlation function data file \texttt{Corr2D.csv}.

In addition to the two-dimensional SANS cross sections, the user can also export the Fourier cross-correlation function components (cross-energy density spectra) of the magnetization vector field (or atomistic magnetic moment lattice). 
These quantities can be optionally exported to the \texttt{SANS2D.csv} file. 
Note that these Fourier correlation functions are generally complex quantities, meaning that the exported data are stored as separate real and imaginary parts:
\begin{align}
    G_{\alpha\beta}^{R}(q,\theta) &= \mathrm{Re}\{ \widetilde{M}_\alpha(q,\theta) \widetilde{M}_{\beta}^{\ast}(q,\theta)\}, \\
    G_{\alpha\beta}^{I}(q,\theta) &= \mathrm{Im}\{ \widetilde{M}_\alpha(q,\theta) \widetilde{M}_{\beta}^{\ast}(q,\theta)\},
\end{align}
where $\alpha,\beta \in \{x, \; y, \; z\}$.

The files \texttt{Corr1D.csv} and \texttt{Corr2D.csv} contain the one-dimensional and two-dimensional correlation functions, respectively. The correlation distance $r$ is given in nm, the correlation functions are expressed in units of nm$^{-4}$, and the pair-distance distribution functions comes in units of nm$^{-2}$.

\section{Verification and applications}
\label{examples}

As part of the validation and benchmarking of the \texttt{NuMagSANS} software package, it is essential to employ reference problems with known analytical solutions. In Section~\ref{sec:Example1}, we examine the classic case of a spherical nanomagnet with a uniform nuclear scattering-length density and a uniform magnetization. This benchmark is widely used in SANS modeling because it is mathematically tractable, physically well understood, and admits closed-form expressions for the relevant real- and reciprocal-space quantities. It is therefore ideally suited for verifying the correctness of the implemented SANS cross-section calculations, conducting regression tests for future software versions, and comparing \texttt{NuMagSANS} output with that of other scattering codes. In addition, the symmetry and simplicity of this model make it a convenient baseline for exploring extensions such as nonuniform magnetization textures, polydispersity, or particle assemblies. In Section~\ref{sec:Example2}, we consider a more complex yet analytically tractable system: a spherical nanomagnet exhibiting a (nonuniform) linear vortex structure. In Section~\ref{sec:Example3}, we demonstrate the integration of \texttt{NuMagSANS} into a \texttt{MuMax3}-based large-scale simulation workflow, using an ensemble of inhomogeneously magnetized iron nanoparticles with randomly oriented magnetic anisotropy axes as a representative test case. Finally, in Section~\ref{sec:Example4}, we assess the computational performance and scaling behavior of \texttt{NuMagSANS} by performing a dedicated runtime experiment using an analytically well-defined micromagnetic test system.

We emphasize that the analytical formulas for the various SANS cross sections and pair-distance distribution functions that are presented in Sections~\ref{sec:Example1} and \ref{sec:Example2} are used to calculate the numerical values given in the Tables~\ref{tab:sphere_ref} and~\ref{tab:vortex_ref}. These should help the reader to easily verify the corresponding numerical output provided by \texttt{NuMagSANS} (Figs.~\ref{fig5}$-$\ref{fig8}).

\subsection{Spherical form factor model}\label{sec:Example1}

\subsubsection*{Model definition}

In real space, the nuclear scattering-length density $N(\mathbf{r})$ and the magnetization vector field $\mathbf{M}(\mathbf{r})$ are modeled as
\begin{align}
N(\mathbf{r}) &= N_0 \, \Theta\!\left( 1 - \frac{r}{R} \right), \\
\mathbf{M}(\mathbf{r}) &= M_0 \, \mathbf{n} \, \Theta\!\left( 1 - \frac{r}{R} \right),
\label{eq:UniformMagnetizationVectorfield}
\end{align}
where $N_0$ is the nuclear scattering-length density contrast (relative to vacuum), $M_0$ is the saturation magnetization, $\mathbf{n} = \{n_x, \; n_y, \; n_z\}$ is a unit vector specifying the magnetization direction, and the Heaviside function $\Theta$ confines the fields to the sphere of radius $R$.

\subsubsection*{Fourier transforms and form factor}

The Fourier transforms of the nuclear and magnetic components are:
\begin{align}
\widetilde{N}(qR) &= \frac{3 V_{\mathrm{s}} N_0 }{(2\pi)^{3/2}} \frac{j_1(q R)}{qR}, \label{eq:SphericalFormFactorMagnetic} \\
\widetilde{\mathbf{M}}(qR) &= \frac{3 V_{\mathrm{s}} M_0 }{(2\pi)^{3/2}} \frac{j_1(q R)}{qR} \, \mathbf{n}, \label{eq:SphericalFormFactorNuclear}
\end{align}
where $V_{\mathrm{s}} = \frac{4\pi}{3} R^3$ and the first-order spherical Bessel function is given by ($\upsilon=qR$)
\begin{align}
j_1(\upsilon) = \frac{\sin \upsilon}{\upsilon^2} - \frac{\cos \upsilon}{\upsilon}.
\end{align}
It is convenient to introduce the normalized sphere form factor
\begin{align}
g(\upsilon) =  \frac{j_1(\upsilon)}{\upsilon} , \quad \lim_{\upsilon \to 0} g(\upsilon) = \frac{1}{3} .
\label{gdef}
\end{align}

\subsubsection*{Differential SANS cross sections ($\mathbf{B}_0 \parallel \mathbf{e}_z \perp \mathbf{k}_0$)}

In the perpendicular scattering geometry the basic SANS cross sections are as follows~\cite{michelsbook}:
\begin{align}
\frac{d\Sigma_{\mathrm{N}}}{d\Omega}(q, \theta) &= 9 N_0^2 V_{\mathrm{s}} [g(qR)]^2, \\
\frac{d\Sigma_{\mathrm{M}}}{d\Omega}(q, \theta) &= 9 b_{\mathrm{H}}^2 M_0^2 V_{\mathrm{s}} [g(qR)]^2 \left(n_x^2 + n_y^2 \cos^2\theta + n_z^2 \sin^2\theta - 2 n_y n_z \sin\theta \cos\theta\right), \\
\frac{d\Sigma_{\mathrm{NM}}}{d\Omega}(q, \theta) &= 18 b_{\mathrm{H}} M_0 N_0 V_{\mathrm{s}} [g(qR)]^2 \left( n_y \sin\theta \cos\theta - n_z \sin^2\theta \right), \\
\frac{d\Sigma_{\mathrm{P}}}{d\Omega}(q,\theta) &= 9 b_{\mathrm{H}}^2 M_0^2 V_{\mathrm{s}} [g(qR)]^2 \left( n_y^2 \sin^2\theta \cos^2\theta + n_z^2 \sin^4\theta - 2 n_y n_z \sin^3\theta \cos\theta \right), \\
\frac{d\Sigma_{\chi}}{d\Omega}(q,\theta) &= 0.
\end{align}

\subsubsection*{Azimuthally averaged SANS cross sections ($\mathbf{B}_0 \parallel \mathbf{e}_z \perp \mathbf{k}_0$)}

The azimuthally averaged SANS cross sections are:
\begin{align}
I_{\mathrm{N}}(q) &= 9 N_0^2 V_{\mathrm{s}} [g(qR)]^2, \\
I_{\mathrm{M}}(q) &= 9 b_{\mathrm{H}}^2 M_0^2 V_{\mathrm{s}} [g(qR)]^2 \frac{2n_x^2 + n_y^2 + n_z^2}{2}, \\
I_{\mathrm{NM}}(q) &= -9 b_{\mathrm{H}} M_0 N_0 V_{\mathrm{s}}[g(qR)]^2 n_z, \\
I_{\mathrm{P}}(q) &= 9 b_{\mathrm{H}}^2 M_0^2 V_{\mathrm{s}} [g(qR)]^2 \frac{n_y^2 + 3 n_z^2}{8}, \\
I_{\chi}(q) &= 0.
\end{align}
For this model, all cross sections scale with the sphere volume $V_{\mathrm{s}}$ due to the definition in equation~\eqref{eq:MagneticSANS} combined with the spherical Fourier transforms equations~\eqref{eq:SphericalFormFactorMagnetic}$-$\eqref{eq:SphericalFormFactorNuclear}.

At $q=0$, the expressions simplify to:
\begin{align}
I_{\mathrm{N}}(0) &=  N_0^2 V_{\mathrm{s}}, \\
I_{\mathrm{M}}(0) &=  b_{\mathrm{H}}^2 M_0^2 V_{\mathrm{s}} \frac{2 n_x^2 + n_y^2 + n_z^2}{2}, \\
I_{\mathrm{NM}}(0) &= -b_{\mathrm{H}} M_0 N_0 V_{\mathrm{s}} n_z, \\
I_{\mathrm{P}}(0) &= b_{\mathrm{H}}^2 M_0^2 V_{\mathrm{s}} \frac{n_y^2 + 3 n_z^2}{8}, \\
I_{\chi}(0) &= 0.
\end{align}

\begin{figure}[!htb]
\centering
\resizebox{0.80\columnwidth}{!}{\includegraphics{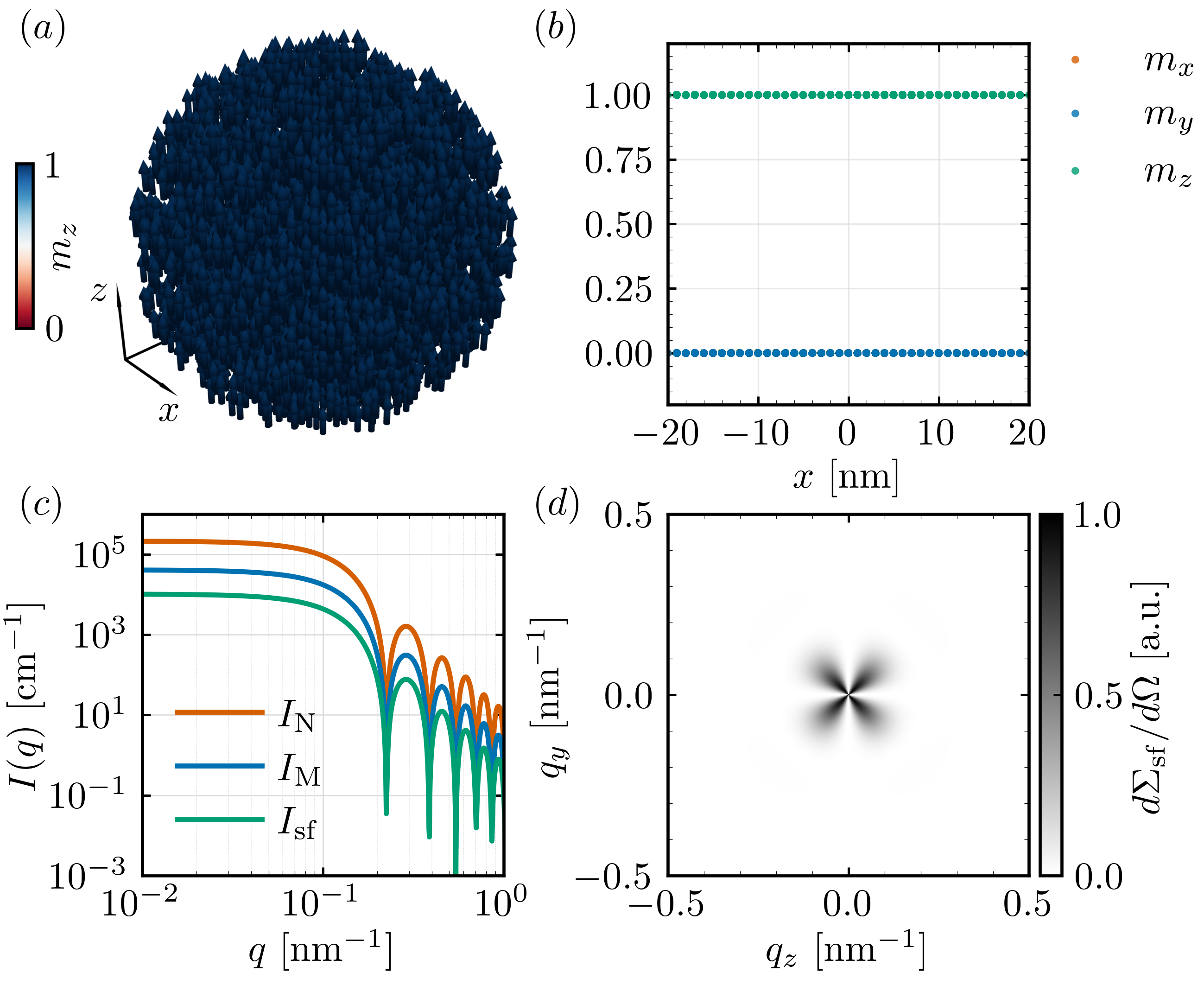}}
\caption{Magnetic SANS from a uniformly magnetized nanoparticle. ($a$)~Three-dimensional visualization of a spherical nanoparticle with a spatially uniform magnetization pointing along the \(z\)~direction, cf.~equation~\eqref{eq:UniformMagnetizationVectorfield}. Colors indicate the \(m_z\)~component. ($b$)~Magnetization profile along the \(x\)~axis through the particle center, showing the expected constant values \(m_x = 0\), \(m_y = 0\), and \(m_z = 1\). ($c$)~Azimuthally averaged SANS cross sections for the same particle (log-log scale): nuclear scattering \(I_{\mathrm{N}}(q)\), unpolarized magnetic scattering \(I_{\mathrm{M}}(q)\), and spin-flip scattering \(I_{\mathrm{sf}}(q)\), cf.~equation~\eqref{eq:iqazi}. The curves exhibit the characteristic form-factor oscillations of a homogeneous sphere. ($d$)~Two-dimensional spin-flip SANS cross section \(d\Sigma_{\mathrm{sf}}/d\Omega(q_y,q_z)\), cf.~equation~\eqref{eq:sfdef}. The intensity shows the expected \(\sin^2\theta \cos^2\theta\) angular dependence for a uniform magnetization
aligned perpendicular to the neutron-beam direction (here: along the \(z\)~axis, with the beam along \(x\)), producing the characteristic four-fold symmetry of the spin-flip scattering pattern.}
\label{fig5}
\end{figure}

\subsubsection*{Pair-distance distribution functions ($\mathbf{B}_0 \parallel \mathbf{e}_z \perp \mathbf{k}_0$)}

Using equations~(\ref{eq:crrad}) and (\ref{eq:prrad}) and the expression for the normalized (squared) sphere form factor $[g(\upsilon)]^2$ (equation~(\ref{gdef})), we obtain the following expression $f(\xi)$, which describes the functionality of the pair-distance distribution function ($\xi = r / R$): 
\begin{align}
    f(\xi) = \frac{1}{4}\xi^2 \left(\xi - 2\right)^2\left( \xi +  4 \right) , \quad \xi \in [0, 2] .
\end{align}
The maximum of this polynomial function in the interval $\xi\in[0, 2]$ is given by: 
\begin{align}
    \xi_{\mathrm{max}} &= \frac{\sqrt{105}}{5} - 1 \cong 1.0494 ,
    \\
    f(\xi_{\mathrm{max}}) &= \frac{144}{5} - \frac{336 \sqrt{105}}{125} \cong 1.2562
\end{align}
Using $f(\xi = r/R)$, the various pair-distance distribution functions read:
\begin{align}
    p_{\mathrm{N}}(r) &= \frac{3 \pi V_{\mathrm{s}} N_0^2}{8 R} f(r/R) ,
    \\
    p_{\mathrm{M}}(r) &= \frac{3 \pi V_{\mathrm{s}} M_0^2 b_{\mathrm{H}}^2}{8 R} f(r/R) \frac{2n_x^2 + n_y^2 + n_z^2}{2} ,
    \\
    p_{\mathrm{NM}}(r) &= -\frac{3\pi V_{\mathrm{s}} N_0 M_0 b_{\mathrm{H}}}{8 R} f(r/R) n_z ,
    \\
    p_{\mathrm{P}}(r) &= \frac{3 \pi V_{\mathrm{s}} M_0^2 b_{\mathrm{H}}^2}{8 R} f(r/R) \frac{n_y^2 + 3n_z^2}{8} ,
    \\
    p_{\chi}(r) &= 0 .
\end{align}

Note that all components of the unit vector $\mathbf{n}$ are considered, in analogy to the Stoner-Wohlfarth model, where the nanoparticles are fully saturated (single domain) but magnetic anisotropy may cause superspin misalignment with respect to $\mathbf{B}_0 \parallel \hat{\mathbf{P}} \parallel \mathbf{e}_z$. This degree of freedom is also useful for testing the \texttt{NuMagSANS} software, particularly for two-dimensional SANS cross sections, where the anisotropy patterns depend on $\mathbf{n}$~\cite{michaeliucrj2023}.

\begin{figure}[!htb]
\centering
\resizebox{0.80\columnwidth}{!}{\includegraphics{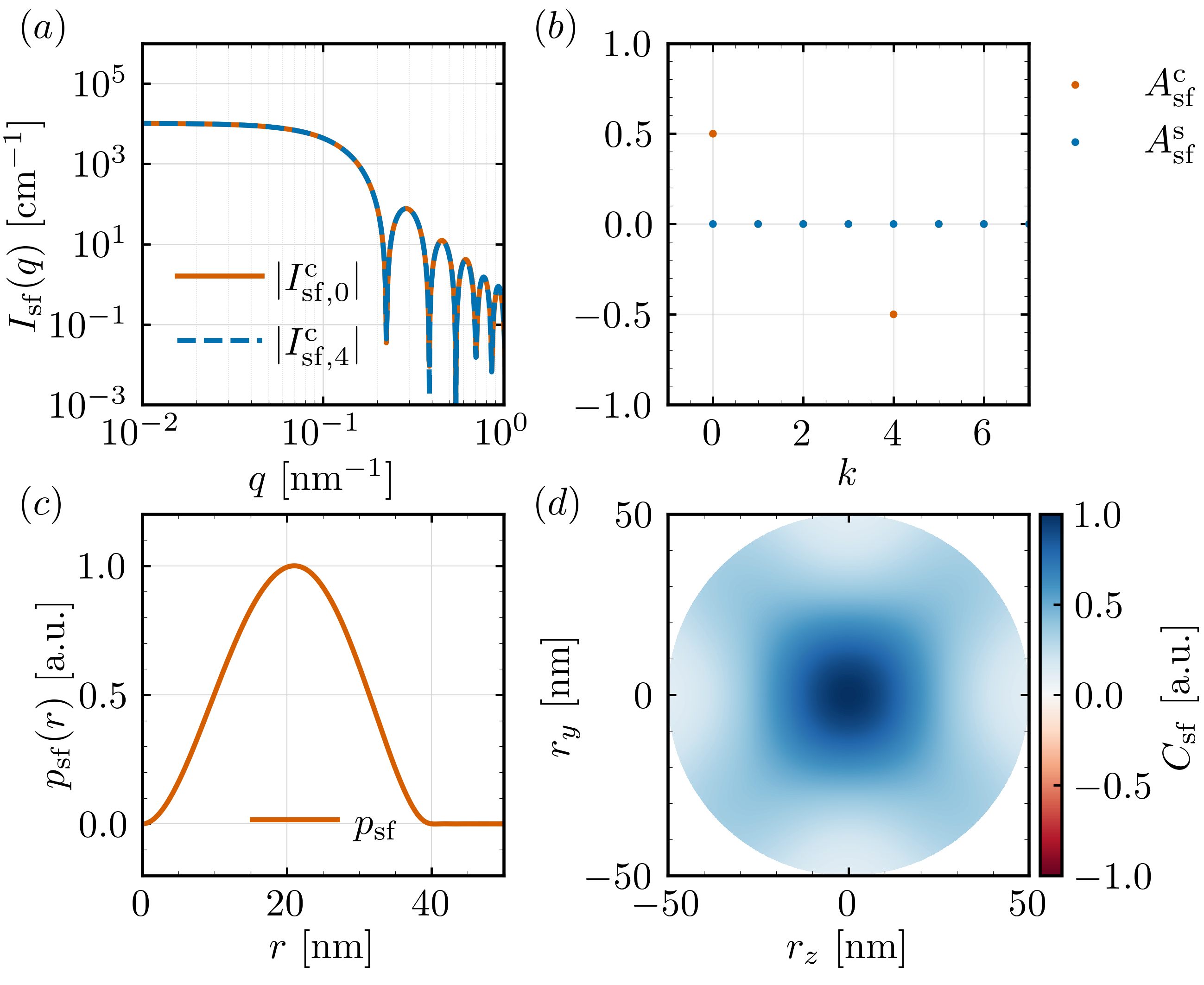}}
\caption{Magnetic SANS from a uniformly magnetized nanoparticle. ($a$)~Absolute cosine-mode intensities of the spin-flip SANS cross section (log-log scale), cf.~equation~\eqref{eq:CosineSinceIntensities}. ($b$)~Modal amplitudes from equation~\eqref{eq:ModalAmplitudes}, highlighting the \(k=4\) contribution that reflects the four-fold symmetry of the two-dimensional spin-flip SANS pattern. ($c$)~Spin-flip pair-distance distribution function \(p_{\mathrm{sf}}(r)\), see equation~\eqref{eq:prrad}. ($d$)~Two-dimensional real-space correlation function \(C_{\mathrm{sf}}(r_y,r_z)\), cf.~equation~\eqref{eq:crav2d}.}
\label{fig6}
\end{figure}

\subsubsection*{Reference example}

For a spherical nanomagnet with a diameter of $D = 40 \, \mathrm{nm}$ ($V_{\mathrm{s}} \cong 3.35103\times 10^{-23} \, \mathrm{m}^3$), magnetized along the $z$~axis ($\mathbf{n} = \mathbf{e}_z$), $M_0 = 1700 \, \mathrm{kA/m}$, and $N_0 = 8.0 \times 10^{14} \, \mathrm{m}^{-2}$, the $q=0$ intensities and the maxima of the pair-distance distribution functions are listed in Table~\ref{tab:sphere_ref}. Figs.~\ref{fig5} and \ref{fig6} display some of the $q$-dependent response functions obtained with \texttt{NuMagSANS}. The data corresponding to this reference example are openly available in the Zenodo repository~\cite{Adams2025_Example1}.

\begin{table}[htb]
\centering
\caption{Analytical reference values for a uniformly magnetized spherical nanomagnet with a diameter of $D = 2R = 40 \, \mathrm{nm}$. Intensities are given at $q=0$; $p_{\mathrm{max}}$ values correspond to $r/R = \xi_{\mathrm{max}}$.}
\label{tab:sphere_ref}
\begin{tabular}{c|c|c|c}
\hline
$I(q=0)$ & Value [cm$^{-1}$] & $p_{\mathrm{max}}$ & Value [nm$^{-2}$] \\
\hline
$I_{\mathrm{N}}(0)$ & $2.145 \times 10^{5}$ & $p_{\mathrm{N}}$ & $15.87\times10^{-4}$ \\
$I_{\mathrm{M}}(0)$ & $0.410 \times 10^{5}$ & $p_{\mathrm{M}}$ & $3.03\times 10^{-4}$ \\
$I_{\mathrm{NM}}(0)$ & $-1.326 \times 10^{5}$ & $p_{\mathrm{NM}}$  & $0$ \\
$I_{\mathrm{P}}(0)$ & $0.308 \times 10^{5}$ & $p_{\mathrm{P}}$ & $2.28\times10^{-4}$ \\
$I_{\chi}(0)$ & $0$ & $p_{\chi}$ & $0$ \\
$I_{\mathrm{sf}}(0)$ & $0.103 \times 10^{5}$ & $p_{\mathrm{sf}}$ & $0.76\times10^{-4}$ \\
$I_{\mathrm{sf}}^{+-}(0)$ & $0.103 \times 10^{5}$ & $p_{\mathrm{sf}}^{+-}$ & $0.76\times10^{-4}$ \\
$I_{\mathrm{sf}}^{-+}(0)$ & $0.103 \times 10^{5}$ & $p_{\mathrm{sf}}^{-+}$ & $0.76\times10^{-4}$ \\
$I_{\mathrm{nsf}}^{++}(0)$ & $1.126 \times 10^{5}$ & $p_{\mathrm{nsf}}^{++}$ & $8.33\times10^{-4}$ \\
$I_{\mathrm{nsf}}^{--}(0)$ & $3.778 \times 10^{5}$ & $p_{\mathrm{nsf}}^{--}$ & $27.96\times10^{-4}$ \\
$I^{+}(0)$ & $1.229 \times 10^{5}$ & $p^{+}$ & $9.09\times10^{-4}$ \\
$I^{-}(0)$ & $3.881 \times 10^{5}$ & $p^{-}$ & $28.72\times10^{-4}$ \\
\hline
\end{tabular}
\end{table}

\subsection{Spherical nanomagnet with linear vortex texture}\label{sec:Example2}

To test the capability of \texttt{NuMagSANS} to handle magnetization textures with broken inversion symmetry and nonzero chirality, it is useful to consider a reference configuration that is still analytically tractable. The spherical nanomagnet with an additional \emph{linear vortex} term represents such a case~\cite{adams2024framework}: the nuclear density remains uniform as in Section~\ref{sec:Example1}, but the magnetization acquires a spatially varying component that mimics the onset of a flux-closure state. This model is valuable for validating the correct treatment of magnetization textures with finite chirality, checking phase factors and imaginary components in the Fourier-space magnetization, and benchmarking the handling of vortex-like states against the uniform magnetization reference.

\subsubsection*{Model definition}

The real-space nuclear density $N(\mathbf{r})$ and magnetization $\mathbf{M}(\mathbf{r})$ are given by \cite{adams2024framework}
\begin{align}
N(\mathbf{r}) &= N_0 \, \Theta\!\left( 1 - \frac{r}{R} \right), \\
\mathbf{M}(\mathbf{r}) &= \left[ M_0 \, \mathbf{e}_z + M_1 \,(x \, \mathbf{e}_y - y \, \mathbf{e}_x) \right] \, \Theta\!\left( 1 - \frac{r}{R} \right),
\label{eq:LinearVortexMagnetizationVectorfield}
\end{align}
where all terms have the same meaning as before. The new parameter $M_1$ models the magnitude of the vortex structure that forms in the $x$-$y$~plane. For a fully saturated state along $\mathbf{e}_z$, $M_1=0$. As the applied magnetic field is reduced towards zero, $M_1$ becomes nonzero, providing a simple phenomenological model for the field dependence of the magnetization texture.

\subsubsection*{Fourier transforms}

Using the Fourier derivative theorem, the $x$ and $y$-dependent terms in $\mathbf{M}(\mathbf{r})$ transform into partial derivatives with respect to $q_x$ and $q_y$:
\begin{align}
\widetilde{N}(\mathbf{q}) &= \frac{3 V_{\mathrm{s}} N_0 }{(2\pi)^{3/2}} \, \frac{j_1(q R)}{qR}, \label{eq:SphericalFormFactorNuclear2} \\
\widetilde{\mathbf{M}}(\mathbf{q}) &= \frac{3 V_{\mathrm{s}}}{(2\pi)^{3/2}} 
\left[ M_0 \, \mathbf{e}_z 
+ i M_1 \left( \mathbf{e}_y \partial_{q_x} - \mathbf{e}_x\partial_{q_y}  \right) \right] 
\frac{j_1(q R)}{qR} .
\label{eq:SphericalFormFactorMagnetic2}
\end{align}
Compared to the uniform case, the Fourier-space magnetization now includes a nonzero imaginary component, directly linked to the broken inversion symmetry introduced by the vortex term. In a linear approximation, such a vector field can model the effects of dipole-dipole coupling, which tends to favor flux-closure-type magnetization textures~\cite{adams2024framework}.

The derivatives in equation~(\ref{eq:SphericalFormFactorMagnetic2}) can be resolved using the chain rule ($\upsilon = qR$):
\begin{align}
    \frac{\partial g(qR)}{\partial q_x}  &= \left.\frac{d g(\upsilon)}{d\upsilon}\right|_{\upsilon = qR} \frac{\partial (qR)}{\partial q_x} 
    = g'(qR)\,\hat{q}_x\, R, \\
    \frac{\partial g(qR)}{\partial q_y}  &= \left.\frac{d g(\upsilon)}{d\upsilon}\right|_{\upsilon = qR} \frac{\partial (qR)}{\partial q_y} 
    = g'(qR)\,\hat{q}_y\, R,
\end{align}
with $\hat{q}_x = q_x/q$, $\hat{q}_y = q_y /q$, and
\begin{align}
    g(\upsilon) &= \frac{j_1(\upsilon)}{\upsilon} = \frac{\sin\upsilon - \upsilon \cos\upsilon}{\upsilon^3},  \quad \lim_{\upsilon \to 0} g(\upsilon) = \frac{1}{3}, \\
    g'(\upsilon) &= \frac{(\upsilon^2 - 3)\sin\upsilon + 3 \upsilon \cos\upsilon}{\upsilon^4},  \quad \lim_{\upsilon \to 0} g'(\upsilon) = 0.
\end{align}

Finally, the Fourier transforms are obtained as:
\begin{align}
\widetilde{N}(\mathbf{q}) &= \frac{3 V_{\mathrm{s}} N_0 }{(2\pi)^{3/2}} \, g(qR), \label{eq:SphericalFormFactorNuclear3}
\\
\widetilde{\mathbf{M}}(\mathbf{q}) &= \frac{3 V_{\mathrm{s}}}{(2\pi)^{3/2}} 
\left[ M_0 \, g(qR)\, \mathbf{e}_z 
+ i M_1 R \, g'(qR) \, \left( \hat{q}_x \mathbf{e}_y - \hat{q}_y \mathbf{e}_x \right) \right] .
\label{eq:SphericalFormFactorMagnetic3}
\end{align}

\begin{figure}[!htb]
\centering
\resizebox{0.80\columnwidth}{!}{\includegraphics{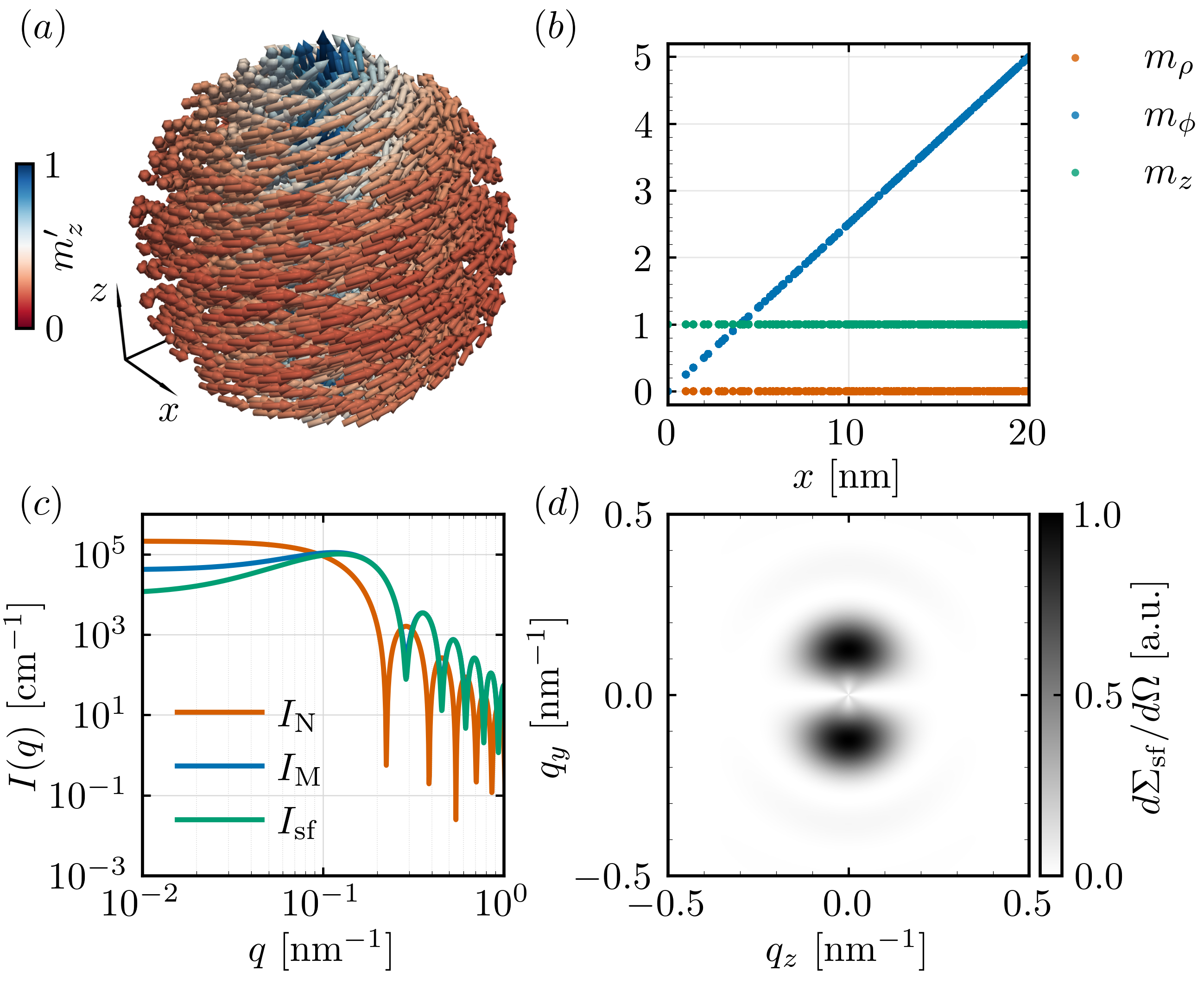}}
\caption{Magnetic SANS from a linear vortex magnetization state. ($a$)~Three-dimensional visualization of the spherical nanoparticle hosting a linear vortex texture. Shown are the normalized magnetization vectors \(\mathbf{m}' = \mathbf{m}/\lVert \mathbf{m} \rVert\), cf.~equation~\eqref{eq:LinearVortexMagnetizationVectorfield}, with colors indicating the normalized axial component \(m_z' = m_z/\lVert \mathbf{m} \rVert\). ($b$)~Magnetization components along the radial coordinate \(\rho\): the expected vortex structure is observed, with \(m_\varphi = 5\rho/R\) increasing linearly with \(\rho\), \(m_\rho = 0\), and a constant axial component \(m_z=1\). ($c$)~Azimuthally averaged SANS cross sections for the same texture (log-log scale): nuclear scattering \(I_{\mathrm{N}}(q)\), unpolarized magnetic scattering \(I_{\mathrm{M}}(q)\), and spin-flip magnetic scattering \(I_{\mathrm{sf}}(q)\), cf.~equation~\eqref{eq:iqazi}. The modified form-factor oscillations reflect the nonuniform, tangential magnetization distribution of the vortex. ($d$)~Two-dimensional spin-flip SANS cross section \(d\Sigma_{\mathrm{sf}}/d\Omega(q_y,q_z)\). The characteristic two-lobe pattern arises from the curling of the vortex magnetization.}
\label{fig7}
\end{figure}

\subsubsection*{Differential SANS cross sections ($\mathbf{B}_0 \parallel \mathbf{e}_z \perp \mathbf{k}_0$)}

For the linear vortex term in the $x$-$y$~plane, the SANS cross sections are:
\begin{align}
    \frac{d\Sigma_{\mathrm{N}}}{d\Omega}(q, \theta) &= 9 N_0^2 V_{\mathrm{s}} [g(qR)]^2, \\
    \frac{d\Sigma_{\mathrm{M}}}{d\Omega}(q, \theta) &= 9 b_{\mathrm{H}}^2  V_{\mathrm{s}} \left(M_0^2[g(qR)]^2 + M_1 R^2 [g'(qR)]^2\right)  \sin^2\theta, \\
    \frac{d\Sigma_{\mathrm{NM}}}{d\Omega}(q, \theta) &= -18  b_{\mathrm{H}} M_0 N_0 V_{\mathrm{s}} [g(qR)]^2  \sin^2\theta, \\
    \frac{d\Sigma_{\mathrm{P}}}{d\Omega}(q,\theta) &= 9 b_{\mathrm{H}}^2 M_0^2 V_{\mathrm{s}} [g(qR)]^2 \sin^4\theta, \\
    \frac{d\Sigma_{\chi}}{d\Omega}(q,\theta) &= 18 V_{\mathrm{s}} b_{\mathrm{H}}^2 M_0 M_1 g(qR) g'(qR) \sin^2\theta \cos\theta.
\end{align}
The angular dependencies differ qualitatively from the uniform case, making this configuration useful for checking the anisotropy handling and the presence of imaginary Fourier components in the software.

\subsubsection*{Azimuthally averaged SANS cross sections ($\mathbf{B}_0 \parallel \mathbf{e}_z \perp \mathbf{k}_0$)}

For unpolarized scattering and $\hat{\mathbf{P}} \parallel \mathbf{e}_z$, the azimuthally averaged cross sections are
\begin{align}
I_{\mathrm{N}}(q) &= 9 N_0^2 V_{\mathrm{s}} [g(qR)]^2, \\
I_{\mathrm{M}}(q) &= \frac{9 V_{\mathrm{s}} b_{\mathrm{H}}^2}{2} \left( M_0^2  [g(qR)]^2 + M_1^2 R^2 [g'(qR)]^2  \right), \\
I_{\mathrm{NM}}(q) &= -9 V_{\mathrm{s}} b_{\mathrm{H}} M_0 N_0  [g(qR)]^2 , \\
I_{\mathrm{P}}(q) &= \frac{27 b_{\mathrm{H}}^2 M_0^2 V_{\mathrm{s}}}{8} [g(qR)]^2, \\
I_{\chi}(q) &= 0.
\end{align}

\begin{figure}[!htb]
\centering
\resizebox{0.80\columnwidth}{!}{\includegraphics{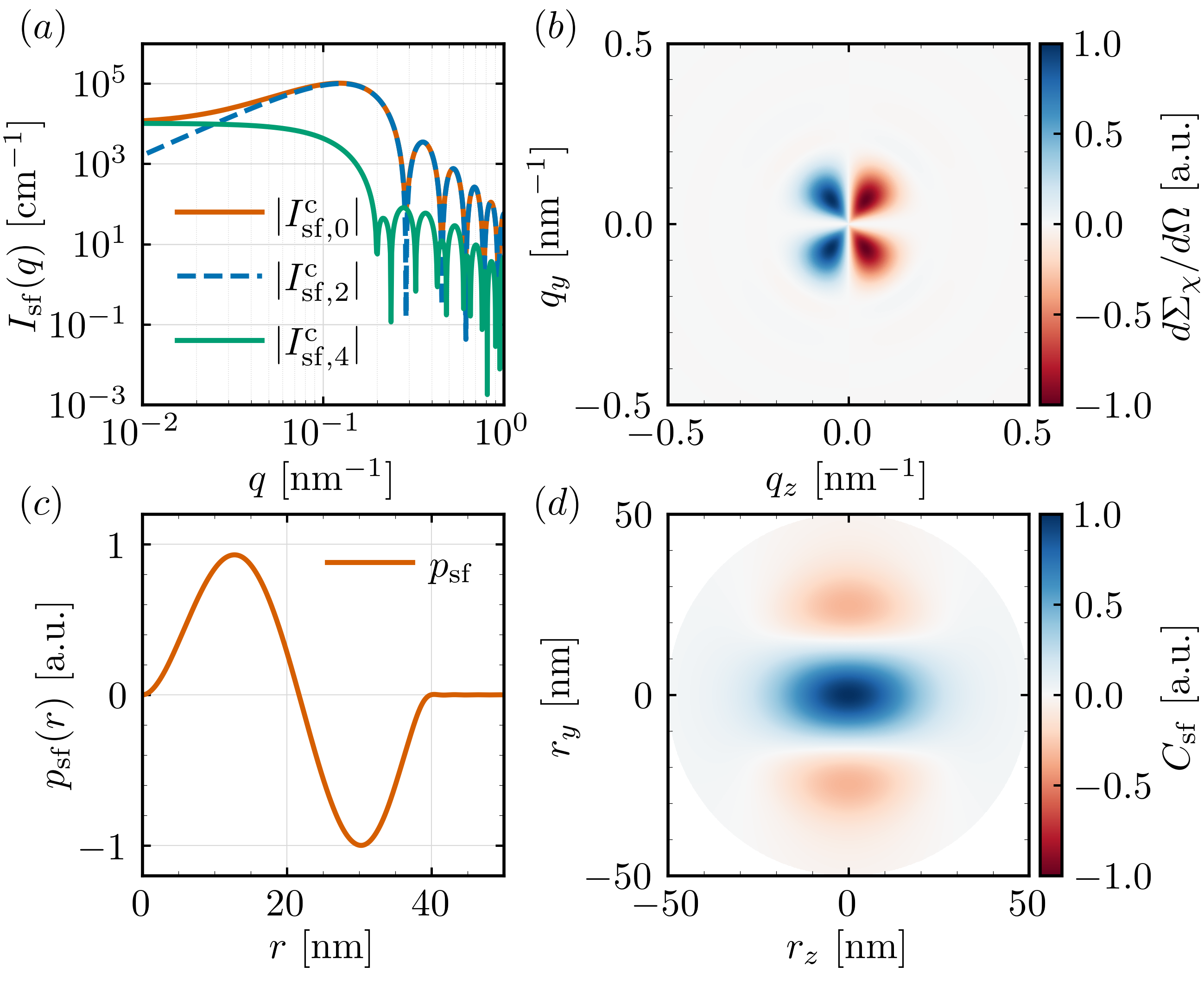}}
\caption{Magnetic SANS from a linear vortex magnetization state. ($a$)~Cosine-mode intensities of the spin-flip SANS cross section \(|I^{c}_{\mathrm{sf},k}(q)|\) (log-log scale), illustrating the dominant contributions for \(k = 0\) and \(k = 2\), which reflect the azimuthal symmetry of the vortex state. Because the linear vortex includes a uniform magnetization component, very small values of \(q\) exhibit a pronounced four-fold angular constellation, whereas at intermediate and larger \(q\) a two-fold anisotropy becomes dominant. ($b$)~Two-dimensional chiral SANS cross section \(d\Sigma_{\chi}/d\Omega(q_y,q_z)\), cf.~equation~\eqref{eq:ChiralSANS}, showing the expected four-lobe antisymmetric pattern associated with the handedness of the vortex. ($c$)~Spin-flip pair-distance distribution function \(p_{\mathrm{sf}}(r)\), demonstrating the oscillatory real-space correlations induced by the curling magnetization. ($d$)~Two-dimensional real-space correlation function \(C_{\mathrm{sf}}(r_y,r_z)\), cf.~equation~\eqref{eq:crav2d}, revealing the characteristic positive-negative correlation structure that reflects the axial and azimuthal components of the linear vortex texture.}
\label{fig8}
\end{figure}

\subsubsection*{Pair-distance distribution functions ($\mathbf{B}_0 \parallel \mathbf{e}_z \perp \mathbf{k}_0$)}

Using the definitions introduced in Section~\ref{sec:Example1}, we derive the analytical pair-distance distribution functions of the linear vortex model. To this end, we introduce the following dimensionless auxiliary functions:
\begin{align}
    f(\xi) &= \frac{1}{4}\,\xi^2 (\xi - 2)^2 (\xi + 4), 
    \qquad \xi \in [0,2], \\
    h(\xi) &= -\frac{1}{32}\,\xi^2 (\xi - 2)^2 \left(\xi^3 + 4\xi^2 + 2\xi - 8\right),
    \qquad \xi \in [0,2].
\end{align}
In terms of these functions, the pair-distance distribution functions associated with the basic SANS observables read:
\begin{align}
    p_{\mathrm{N}}(r) &=
    \frac{3\pi V_{\mathrm{s}} N_0^2}{8R}\, f(r/R), \\
    p_{\mathrm{M}}(r) &=
    \frac{3\pi V_{\mathrm{s}} M_0^2 b_{\mathrm{H}}^2}{16R}\, f(r/R)
    + \frac{9\pi V_{\mathrm{s}} M_1^2 R^2 b_{\mathrm{H}}^2}{20R}\, h(r/R), \\
    p_{\mathrm{NM}}(r) &=
    -\frac{3\pi V_{\mathrm{s}} N_0 M_0 b_{\mathrm{H}}}{8R}\, f(r/R), \\
    p_{\mathrm{P}}(r) &=
    \frac{9\pi V_{\mathrm{s}} M_0^2 b_{\mathrm{H}}^2}{64R}\, f(r/R), \\
    p_{\chi}(r) &= 0.
\end{align}

\subsubsection*{Reference example}

For a sphere with $D = 40\,\mathrm{nm}$ ($V_{\mathrm{s}} \cong 3.35103\times10^{-23} \; \mathrm{m}^3$), $M_0 = 1700 \, \mathrm{kA/m}$, $M_1 = 5M_0/R$, and $N_0 = 8.0 \times 10^{14} \, \mathrm{m}^{-2}$, the $q=0$~intensities and $p_{\mathrm{max}}$ values are summarized in Table~\ref{tab:vortex_ref}. Figs.~\ref{fig7} and \ref{fig8} display some of the $q$-dependent response functions obtained with \texttt{NuMagSANS}. The data corresponding to this reference example are openly available in the Zenodo repository~\cite{Adams2025_Example2}.

\begin{table}[htb]
\centering
\caption{Analytical reference values for a spherical nanomagnet with a linear vortex texture ($D = 40 \, \mathrm{nm}$). Intensities are given at $q=0$; $p_{\mathrm{max}}$ values correspond to $r/R = \xi_{\mathrm{max}}$.}
\label{tab:vortex_ref}
\begin{tabular}{c|c|c|c|c|c}
\hline
$I(q=0)$ & Value [cm$^{-1}$] & $I_{\mathrm{max}}$ & Value [cm$^{-1}$] & $p_{\mathrm{max}}$ &  Value [nm$^{-2}$] \\
\hline
$I_{\mathrm{N}}(0)$ & $2.145 \times 10^{5}$ & 
$I_{\mathrm{N}} $ & $2.145 \times 10^{5}$ & 
$p_{\mathrm{N}}$ & $15.87\times10^{-4}$ \\
$I_{\mathrm{M}}(0)$ & $0.410 \times 10^{5}$ & 
$I_{\mathrm{M}} $ & $1.113\times10^{5}$ & 
$p_{\mathrm{M}}$ &  $18.68\times 10^{-4}$  \\
$I_{\mathrm{NM}}(0)$ & $-1.326 \times 10^{5}$ & 
$I_{\mathrm{NM}} $ & $0$ & 
$p_{\mathrm{NM}}$  &  $0$  \\
$I_{\mathrm{P}}(0)$ & $0.308 \times 10^{5}$ & 
$I_{\mathrm{P}} $ & $0.308 \times 10^{5}$ & 
$p_{\mathrm{P}}$  & $2.28\times10^{-4}$ \\
$I_{\chi}(0)$ & $0$ & 
$I_{\chi}$ & $0$ & 
$p_{\chi}$ & $0$ \\
$I_{\mathrm{sf}}(0)$ & $0.103 \times 10^{5}$ & 
$I_{\mathrm{sf}} $ & $1.025\times10^{5}$ & 
$p_{\mathrm{sf}}$  & $17.07\times10^{-4}$  \\
$I_{\mathrm{sf}}^{+-}(0)$ & $0.103 \times 10^{5}$ & 
$I_{\mathrm{sf}}^{+-}$ & $1.025\times10^{5}$ & 
$p_{\mathrm{sf}}^{+-}$  & $17.07\times10^{-4}$   \\
$I_{\mathrm{sf}}^{-+}(0)$ & $0.103 \times 10^{5}$ & 
$I_{\mathrm{sf}}^{-+}$ & $1.025\times10^{5}$ & 
$p_{\mathrm{sf}}^{-+}$  & $17.07\times10^{-4}$  \\
$I_{\mathrm{nsf}}^{++}(0)$ & $1.126 \times 10^{5}$ & 
$I_{\mathrm{nsf}}^{++}$ & $1.126 \times 10^{5}$ & 
$p_{\mathrm{nsf}}^{++}$  & $8.33\times10^{-4}$   \\
$I_{\mathrm{nsf}}^{--}(0)$ & $3.778 \times 10^{5}$ & 
$I_{\mathrm{nsf}}^{--}$ & $3.778 \times 10^{5}$ & 
$p_{\mathrm{nsf}}^{--}$ & $27.96\times10^{-4}$  \\
$I^{+}(0)$ & $1.229 \times 10^{5}$ & 
$I^{+}$ & $1.448 \times 10^{5}$ & 
$p^{+}$ & $23.17\times10^{-4}$  \\
$I^{-}(0)$ & $3.881 \times 10^{5}$ & 
$I^{-}$ & $3.881 \times 10^{5}$ & 
$p^{-}$ & $39.22\times10^{-4}$  \\
\hline
\end{tabular}
\end{table}

\subsection{Spherical iron nanoparticles:~effect of interparticle interference}~\label{sec:Example3}

This example expands on previous work~\cite{evelynprb2023,sinaga2024neutron,adams2024framework} that combines \texttt{NuMagSANS} with \texttt{MuMax3} in a micromagnetic SANS workflow. The previous studies focused on dilute, noninteracting ensembles of iron nanoparticles with dipolar-energy-driven internal vortex structures inside of each particle, i.e., the magnetodipolar coupling {\it between} particles has been neglected. Here, as an example for interparticle interference scattering, we consider positional correlations between the particles while still neglecting the interparticle dipole-dipole coupling.

The simulations involve 800 spherical iron nanoparticles arranged in three different configurations: a dilute ensemble, an isotropic random arrangement, and an ordered simple cubic lattice structure. The simulations are conducted for the remanent state (zero applied field after prior saturation). By including structural interparticle interference, this study captures the collective effects often observed in experimental systems.

\begin{figure}[htb!]
\centering
\resizebox{0.85\columnwidth}{!}{\includegraphics{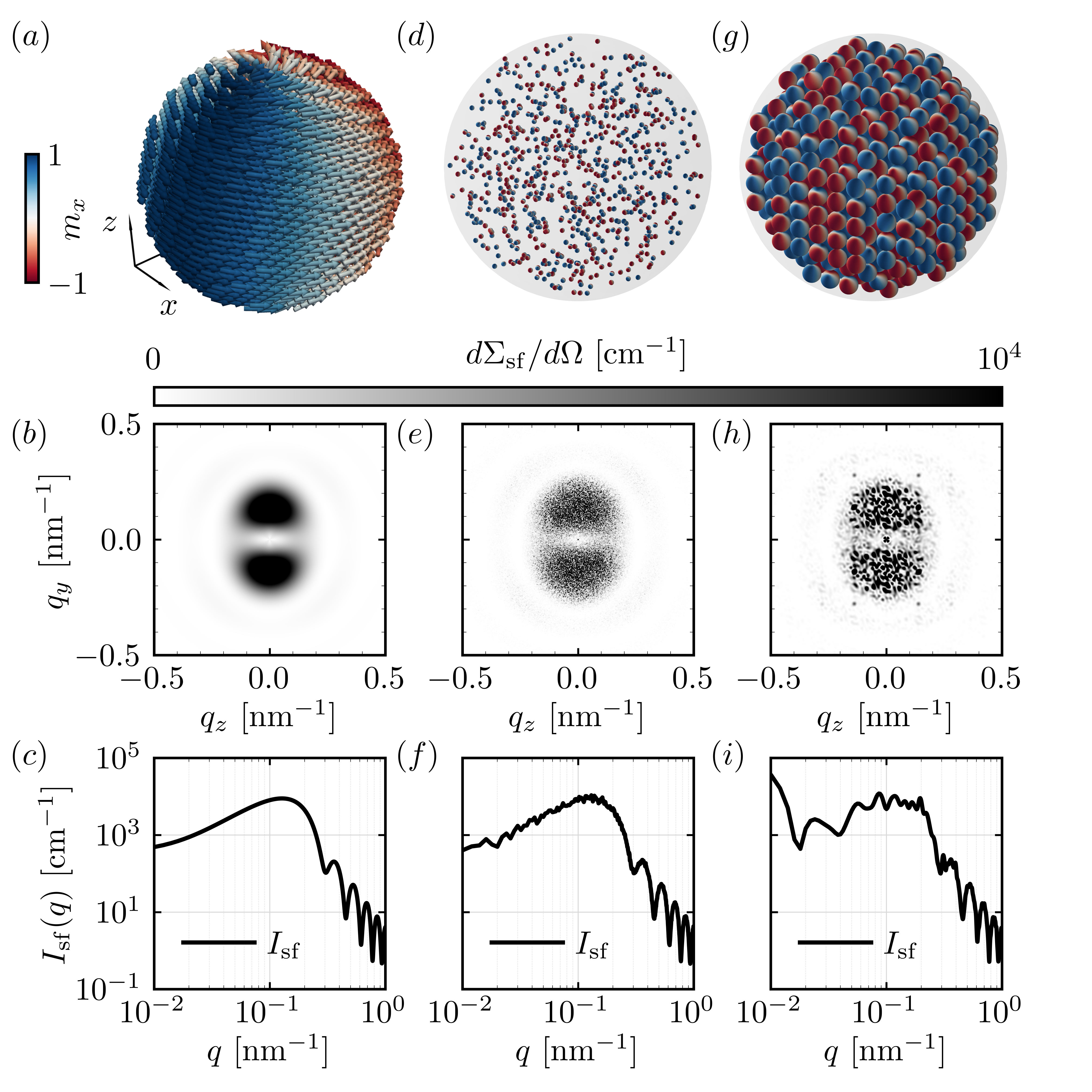}}
\caption{Spin-flip SANS observables at the remanent state for three different ensembles of 800 iron nanoparticles with equal diameters of \(D = 40\,\mathrm{nm}\) (micromagnetic spin textures were simulated with \texttt{MuMax3}). Left column ($a$, $b$, $c$): dilute ensemble (single particle case). ($a$)~Representative particle with its internal spin texture color-coded by \(m_x\). ($b$)~Corresponding two-dimensional spin-flip SANS cross section \(d\Sigma_{\mathrm{sf}}/d\Omega(q_y,q_z)\) showing a smooth intensity distribution that is characteristic of weak interparticle correlations. ($c$)~Azimuthally averaged spin-flip intensity \(I_{\mathrm{sf}}(q)\) displaying the expected sphere-like form-factor oscillations (log-log scale). Middle column ($d$, $e$, $f$): moderately dense-packed system. ($d$)~Particles are positioned inside a supersphere with a volume fraction of \(\eta = 5\,\%\). The two-dimensional~($e$) and the one-dimensional~($f$) spin-flip SANS cross sections exhibit an increased statistical noise due to the disordered arrangement of the particles. Right column ($g$, $h$, $i$): simple cubic structure. ($g$)~Particles are placed on a simple cubic lattice inside a supersphere (volume fraction: $\eta = 30 \, \%$; center-to-center distance between spheres: $45 \, \mathrm{nm}$). ($h$)~The two-dimensional spin-flip SANS cross section now shows emerging Braggpeak-like features reflecting the underlying periodic packing. ($i$)~\(I_{\mathrm{sf}}(q)\) displays modulations and peak structures associated with these lattice correlations.}
\label{fig9}
\end{figure}

Fig.~\ref{fig9} displays the spatial arrangement of the nanoparticles (top row) and their corresponding two-dimensional (middle row) and one-dimensional (lower row) spin-flip SANS cross sections. Unlike the smooth SANS signal observed for the dilute case (left column), the middle and right columns exhibit noisy SANS signals due to the included interparticle interference scattering. Additionally, Fig.~\ref{fig9}($f$) shows Braggpeak-like features due to simple cubic ordering.

As a reference, the execution time for a single system configuration in this example (with only one applied magnetic field value) is approximately $60 \, \mathrm{s}$ on an NVIDIA RTX~3090~GPU. The simulation comprises a total of $\sim$$3.38 \times 10^6$ magnetization cells, corresponding to $800$ nanoparticles with $4224$ cells per particle. A uniform nuclear scattering length is assumed, nuclear data files are included, and all the possible \texttt{NuMagSANS} output options are enabled. The reciprocal-space sampling is performed on a $\{q,\theta\}$~grid of $1000 \times 1000$ points, and the corresponding real-space correlation functions are also evaluated on a $\{\rho,\alpha\}$~grid of $1000 \times 1000$ points. The full output of this single system configuration amounts to approximately $670 \, \mathrm{MB}$ in CSV~format, reflecting the high reciprocal-space and real-space sampling resolutions.

In many practical cases, not all the available output quantities are required (e.g., the two-dimensional correlation functions). This can substantially reduce the execution time of \texttt{NuMagSANS}. Under such conditions, parameter sweeps along a hysteresis curve comprising applied field values of the order of $1000$ become computationally feasible within a few hours on a single GPU.

The data corresponding to this reference example are openly available in the Zenodo repository~\cite{Adams2025_Example3}.

\subsection{Runtime experiment}
\label{sec:Example4}

To assess the computational performance and scaling behavior of \texttt{NuMagSANS}, we performed a dedicated runtime experiment based on a controlled and analytically well-defined micromagnetic test system. Specifically, we consider a linear vortex-like magnetization configuration,
\begin{align}
    \mathbf{m}(\mathbf{r}) =
    \frac{\left\{-5y/R,\; +5x/R,\; 1\right\}}
    {\sqrt{(5x/R)^2 + (5y/R)^2 + 1}} ,
\end{align}
defined inside a spherical nanoparticle of diameter $D = 2R = 40\, \mathrm{nm}$. The nuclear scattering-length density is assumed to be spatially uniform.

The nanoparticle is discretized on a simple cubic lattice with a cell size of $a = 2 \,\mathrm{nm}$, resulting in a total of $4169$ cells per particle. This configuration is representative of the micromagnetic vortex states considered throughout this work, while remaining sufficiently simple to isolate computational scaling effects from physical complexity.

\begin{figure}[htb!]
\centering
\resizebox{0.50\columnwidth}{!}{\includegraphics{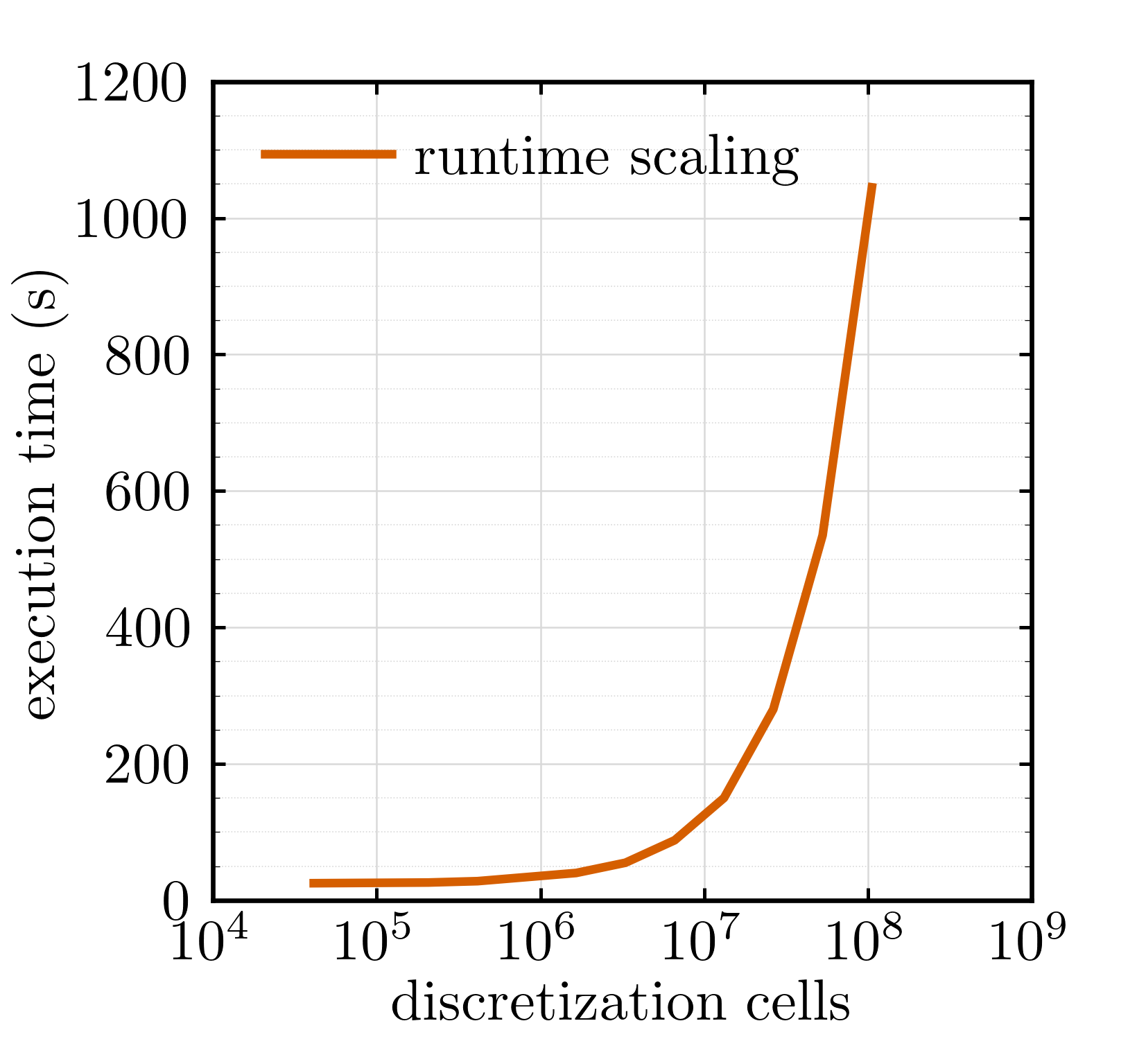}}
\caption{Execution time of the \texttt{NuMagSANS} software as a function of the total number of discretization cells, measured on a single NVIDIA RTX~3090~GPU (log-linear scale).}
\label{fig10}
\end{figure}

The runtime benchmark is carried out by increasing the number of identical nanoparticles from $N = 10$ (corresponding to $41\,690$ cells) up to $N = 25\,600$ (corresponding to approximately $1.067 \times 10^8$ cells). All simulations are executed on a single NVIDIA RTX~3090~GPU using identical numerical settings and output options.

Figure~\ref{fig10} summarizes the resulting execution times as a function of the total number of discretization cells. For small system sizes, the run time is dominated by a nearly constant overhead associated with initialization and data setup. Beyond this regime, the execution time increases approximately linearly with the number of cells over more than two orders of magnitude in system size. At the largest system sizes, a moderate increase in slope is observed, which can be attributed to increased GPU memory traffic and reduced cache efficiency. Importantly, no superlinear scaling behavior or numerical instabilities are observed within the investigated range.

These results demonstrate that \texttt{NuMagSANS} enables the efficient evaluation of large-scale micromagnetic SANS problems comprising tens of millions of discretization cells on a single GPU, thereby bridging the gap between desktop-scale simulations and traditional large-scale high-performance computing approaches.

The data corresponding to this reference example are openly available in the Zenodo repository~\cite{Adams2026_Example4}.

\section{Summary and conclusion}

We have presented \texttt{NuMagSANS}, a GPU-accelerated framework for calculating nuclear and magnetic small-angle neutron scattering (SANS) cross sections and correlation functions of complex systems. The software accounts for nuclear, magnetic, and interference scattering contributions and enables the computation of unpolarized, polarized, chiral, and azimuthally averaged observables, facilitating the direct comparison between theoretical models and experimental SANS measurements.

The modular design of \texttt{NuMagSANS} allows users to import user-defined nuclear and magnetic real-space data, making it a versatile tool for investigating a wide range of systems, including bulk magnetic materials, magnetic nanoparticles, and ordered assemblies of nanoscaled objects. Its GPU-accelerated implementation ensures high computational efficiency, enabling simulations of large-scale systems. The versatility of \texttt{NuMagSANS} has been shown through applications to dilute as well as densely-packed arrangements of spherical iron nanoparticles. A particular strength of \texttt{NuMagSANS} lies in the possibility of seamlessly integrating the results of large-scale micromagnetic simulations (performed, e.g.\ with \texttt{MuMax3}) into a SANS workflow.

Looking ahead, \texttt{NuMagSANS} offers numerous opportunities for further development. Future work could include the extension of the software to account for resolution effects and the inclusion of time-dependent magnetic scattering phenomena. These developments would further expand its applicability and enhance its relevance to the SANS community.

In conclusion, \texttt{NuMagSANS} provides a powerful and flexible platform for advancing the understanding of nuclear and magnetic SANS phenomena. Its ability to simulate complex systems and incorporate collective effects makes it a valuable tool for researchers studying nanoscale magnetic and structural interactions.

\subsection*{Acknowledgments}
We thank Evelyn Pratami Sinaga, Ivan Titov, \v{S}tefan Li\v{s}\v{c}\'{a}k, Elizabeth Jefremovas, and Annika Stellhorn for fruitful discussions. 
We acknowledge financial support from the National Research Fund of Luxembourg (AFR Grant No.~15639149).

\subsection*{Data Availability}
The data sets generated and analyzed during the current study are available in the Zenodo repositories \cite{Adams2025_Example1, Adams2025_Example2, Adams2025_Example3,Adams2026_Example4}. The open-source simulation and analysis software \texttt{NuMagSANS} is available under the MIT License at \url{https://github.com/AdamsMP92/NuMagSANS.git}.

\bibliography{RefNew,BIB}

\end{document}